\documentclass[final,5p,times,twocolumn]{elsarticle}

\usepackage{amssymb}
\usepackage{caption}
\usepackage{subcaption}
\usepackage{mwe}
\usepackage{amsmath}
\usepackage{color}
\usepackage{hyperref}
\journal{Elsevier}

\begin{document}

\begin{frontmatter}



\title{A Systematic Assessment of Data Volume Reduction for IACTs}
%%% A novel approach - or a systematic assessment? Jim


\author[label1]{Clara Escañuela Nieves\corref{cor1}}
\cortext[cor1]{Corresponding author}
\ead{clara.escanuela@mpi-hd.mpg.de}
\author[label1]{Felix Werner}
\author[label1]{Jim Hinton}

\affiliation[label1]{organization={Max-Planck-Institut für Kernphysik},%Department and Organization
            addressline={Saupfercheckweg 1}, 
            city={Heidelberg},
            postcode={69126}, 
            country={Germany}}

\begin{abstract}

High-energy cosmic rays generate air showers of secondary particles when they enter the Earth's atmosphere. These highly energetic particles emit Cherenkov light that can be detected by Imaging Air Cherenkov Telescopes~(IACTs) or Water-Cherenkov Detectors at mountain altitudes. Advances in the technique and larger collection areas have increased the rate at which air shower events can be captured, and the amount of data produced by modern high-time-resolution Cherenkov cameras. Therefore, {\it Data Volume Reduction}~(DVR) has become critical for such telescope arrays, ensuring that only relevant information regarding the air shower is stored long-term. Given the vast amount of raw data, owing to the highest resolution and sensitivity, the upcoming Cherenkov Telescope Array Observatory~(CTAO) will need robust data reduction strategies to ensure efficient data handling and a sustainable data analysis. The CTAO data rates needs to be reduced from hundreds of Petabytes~(PB) per year to a few PB/year.

This paper presents DVR algorithms tailored for CTAO but also applicable for other existing IACT arrays, focusing on the selection of pixels likely to contain Cherenkov light from the air shower. It describes and evaluates multiple algorithms based on their signal efficiency, noise rejection, and shower reconstruction. With a focus on a time-based clustering algorithm which demonstrates a notable enhancement in the retention of low level signal pixels. Moreover, the robustness is assessed under different observing conditions, including detector defects. Through testing and analysis, it is shown that these algorithms offer promising solutions for efficient volume reduction in CTAO. They effectively address the challenges posed by the array's very large data volume and ensure reliable data storage amidst varying observational conditions and hardware issues.

\end{abstract}


\end{frontmatter}


\section{Introduction}\label{sec:intro}

High-energy gamma rays that enter Earth's atmosphere interact with air molecules, triggering a cascade of secondary particles, which in turn emit flashes of Cherenkov light~\citep{technique}. This Cherenkov light is detected by arrays of telescopes equipped with high-speed cameras capable of nanosecond~(ns) time sampling. These cameras, typically consisting of thousands of light-sensitive detectors, capture the faint and rapid flashes of Cherenkov radiation. The resulting images on the cameras usually manifest as elliptical patterns. By analysing the shape and characteristics of these images, scientists can determine both the energy and direction of the incoming gamma ray. However, proton-induced showers, which occur more frequently, can also trigger Cherenkov cameras. Fortunately, proton showers leave distinct patterns in the data, allowing researchers to distinguish them from gamma rays and improve the accuracy of gamma-ray detection.

The forthcoming CTAO~\citep{ctao} will be the next-generation gamma-ray detector of its kind. It will have two observation sites: one in Paranal, Chile, and the other in La Palma, Spain. The northern hemisphere array will consist of four Large-Sized Telescopes~(LSTs) and nine Medium-Sized Telescopes~(MSTs)~\citep{website_layout}, with a focus on low to mid-energy~(from 20\,GeV to 5\,TeV) gamma rays due to its limited effective area. The southern array, located in Paranal, will feature 51 telescopes, including 14 MSTs and 37 Small-Sized Telescopes~(SSTs). With its wide energy range~(150\,GeV~-~300\,TeV) and view of the inner galaxy, the southern site is well-suited for Galactic observations, while also prioritising extra-galactic sources. Future enhancements may also include LSTs in the South.

The CTAO will, therefore, feature an increased number of telescopes and incorporate wider field-of-view cameras with enhanced data-cube capture, significantly improving performance compared to the current generation of IACT arrays. However, this advancement presents a major challenge: managing the vast data rates generated by its network of telescopes. The recorded telescope-wise data consists of pixel-wise waveform data per each triggered event, resulting in an enormous influx of data. To maintain operational efficiency, a crucial aspect will be the implementation of significant data reduction strategies, ensuring that storage requirements are met while maintaining scientific output.

Outbound bandwidth, processing resources and long-term storage are limited. For cost-effective operation, DVR is expected to reduce the amount of data from a factor of 10 in early operation up to $\sim$50 once complete. An extra factor of 2 is achieved by lossless compression to limit the data rate from hundreds of PB per year to $\sim$3\,PB/y at each site~\citep{website}. 

Data reduction in the CTAO involves two key steps based on per-pixel information. First, data can significantly be reduced by selecting only the pixels that contain meaningful information for event reconstruction. Noise pixels, which likely carry no signal from the original shower, can be discarded. This paper presents various algorithms designed to achieve this. An ideal method would accurately identify and store all signal pixels for offline analysis, leading to substantial data reduction while ensuring the detector functions effectively.

Furthermore, data reduction can be enhanced by shortening the traces~—~signals generated by a light detector in response to photon illumination. Several tens of ns after triggering, bins with no signal could be ignored. The trace~(typically greater than 100\,ns) length is chosen to accommodate bright, large-impact-distance showers, with long signal pulses to capture the extended pulse tail. Consequently, the trace can often be shortened on a per-pixel basis, particularly for high-rate, low-energy events.

DVR algorithms are verified based on the signal efficiency and noise false positive rate, with the objective to maximise the signal efficiency and data reduction. As for DVR, the robustness and reliability of the methods must be taken into account. We evaluate the performance of the methods under increasing Night Sky Background~(NSB), broken pixels, and calibration uncertainty. 

The methods described in the paper cannot only be applied for real-time data reduction, but also, for offline cleaning of the images to remove noise. We test the performance of the methods for cleaning by comparing the high-level performance of the telescope array. Cleaning methods are developed to improve the sensitivity of Cherenkov detectors with better angular and energy resolution and with a focus on gamma-hadron separation. 

We describe multiple DVR algorithms for the CTAO, applicable to other observatories, in Section~\ref{sec:methods}. In Section~\ref{sec:verification}, the performance of each method is based on the signal kept and the DVR factor achieved. As well as the angular and energy resolution for multiple levels of data reduction. The verification is further extended in Section~\ref{sec:robustness} where we explore the robustness of the methods. In the previous mentioned sections, methods are presented and tested with simulations of MSTs in the South with FlashCam~\citep{flashcam}. Section~\ref{sec:cameras} discusses the performance of the algorithms when applied to all CTAO Cherenkov cameras. Lastly, we discuss the results.

\begin{figure}[h!]
    \centering
    \includegraphics[width=\linewidth]{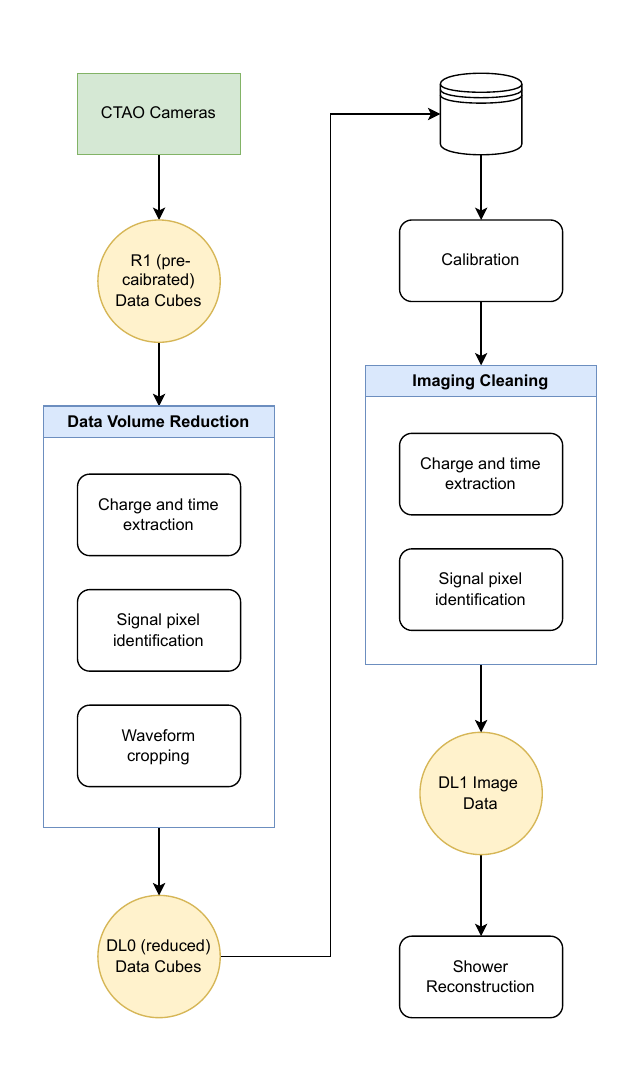}
    \caption{Illustration of the CTAO data flow~\citep{data_model_ref} from on-site/real-time activities~(left side) to offline processing up to the level of air shower reconstruction~(right side), illustrating the role of the closely related activities of {\it Data Volume Reduction} and {\it Image Cleaning}. R1/DL0 and DL1 are data levels defined in the CTAO Data model. R1 differs from the true camera raw data in being pre-calibrated and converted to a common format. Data Level 0~(DL0) is the lowest level data that is transferred from the observatory sites and preserved long-term.}
    \label{fig:sketch}
\end{figure}

\section{Methods}\label{sec:methods}

Fig.~\ref{fig:sketch} illustrates the role of {\it Data Volume Reduction} in the CTAO data processing scheme. After the calibration of R1 traces, the charge and peak arrival time~(image) of waveforms are extracted before DVR. Two key reduction factors are considered: signal pixel selection and waveform cropping. Signal pixels are identified from the extracted image based on the probability that each pixel contains a signal from the shower. Following this, R1 traces from the surviving pixels identified as signal can be cropped to eliminate time samples without signal, ensuring that the entire signal pulse remains intact.

The traces that survive the two-step reduction process~(referred to as DL0) are stored for a more careful recomputation of the charge and time information, incorporating improved calibration. This is followed by a signal pixel identification step, which may resemble the approach used during the DVR process. The charge extraction and pixel selection procedures constitute the {\it Image Cleaning} phase, yielding the DL1 image data.

After the {\it Image Cleaning} process, events are parameterised, and the images are characterised by a set of parameters, such as Hillas parameters~\citep{hillas}. These parameters, along with additional relevant information, are used as inputs to train machine learning algorithms for shower reconstruction~(as seen in~\citep{freepact}) and background rejection. This process reconstructs the nature of the initiating particle with a certain probability and estimates the energy and direction of the incoming primary particle, assuming it is a gamma ray. The effectiveness of these methods is tested through simulations, comparing the proximity between the reconstructed and true parameters of the simulated shower.

\begin{figure}
    \centering
    \includegraphics[width=\linewidth]{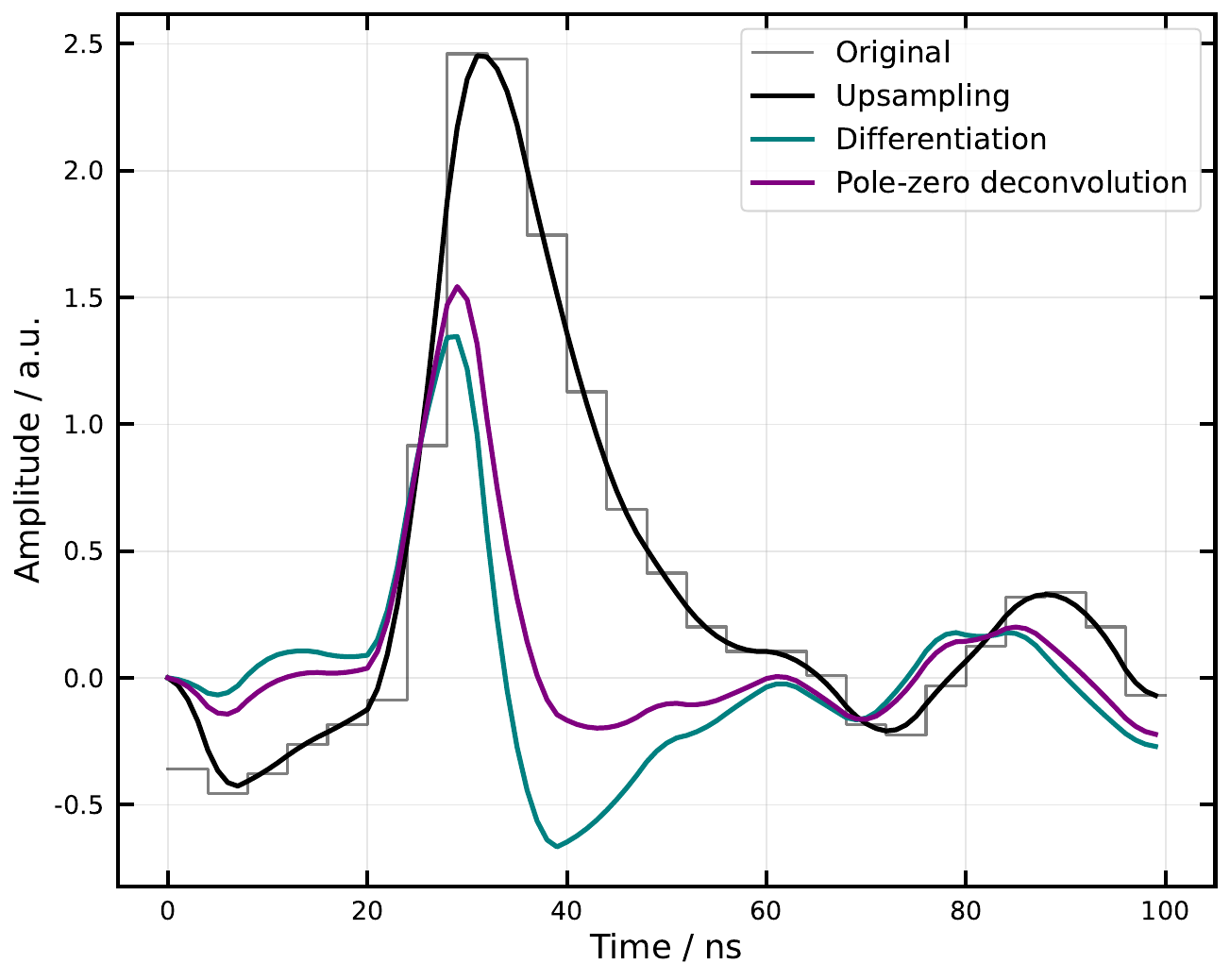}
    \caption{Waveform from a simulated MST equipped with a FlashCam, for a specific pixel and event. The calibrated waveform, after pedestal subtraction and gain calibration, is shown together with the results of three processing steps: upsampling, differentiation and pole-zero deconvolution. }
    \label{fig:wfs_pz}
\end{figure}

As mentioned earlier, {\it Data Volume Reduction} relies on accurately extracting the number of photo-electrons~(PEs) at each pixel and their arrival times. The image extraction for each pixel is based on information from its nearest neighbouring pixels. FlashCam samples waveforms at a rate of 250\,MHz, which is about four times lower than those used by other Cherenkov cameras. To improve accuracy, linear interpolation and smoothing are applied to the waveforms, as illustrated in Fig.~\ref{fig:wfs_pz}~(Upsampling). Following this, pole-zero deconvolution~\citep{pz_dec} is applied to the upsampled waveforms to address the long pulse tail. The deconvolved traces are then integrated around the peak position, determined as the average of the maxima of the nearest neighbours' traces. The integration window spans 7\,ns, with 3\,ns summed to the left of the peak and 4\,ns to the right. The signal timing is obtained from the maximum of the differentiated trace, corresponding to the leading edge. The result of the integral is then adjusted for gain loss after deconvolution, and a time shift correction is applied. Fig.~\ref{fig:wfs_pz}~(Deconvolution and differentiation) illustrates these steps before any gain or time shift correction. The differentiated trace initially shows a clear undershoot, which is flattened after deconvolution.

\begin{figure}[h!]
\begin{subfigure}{\linewidth}
\includegraphics[width=\linewidth]{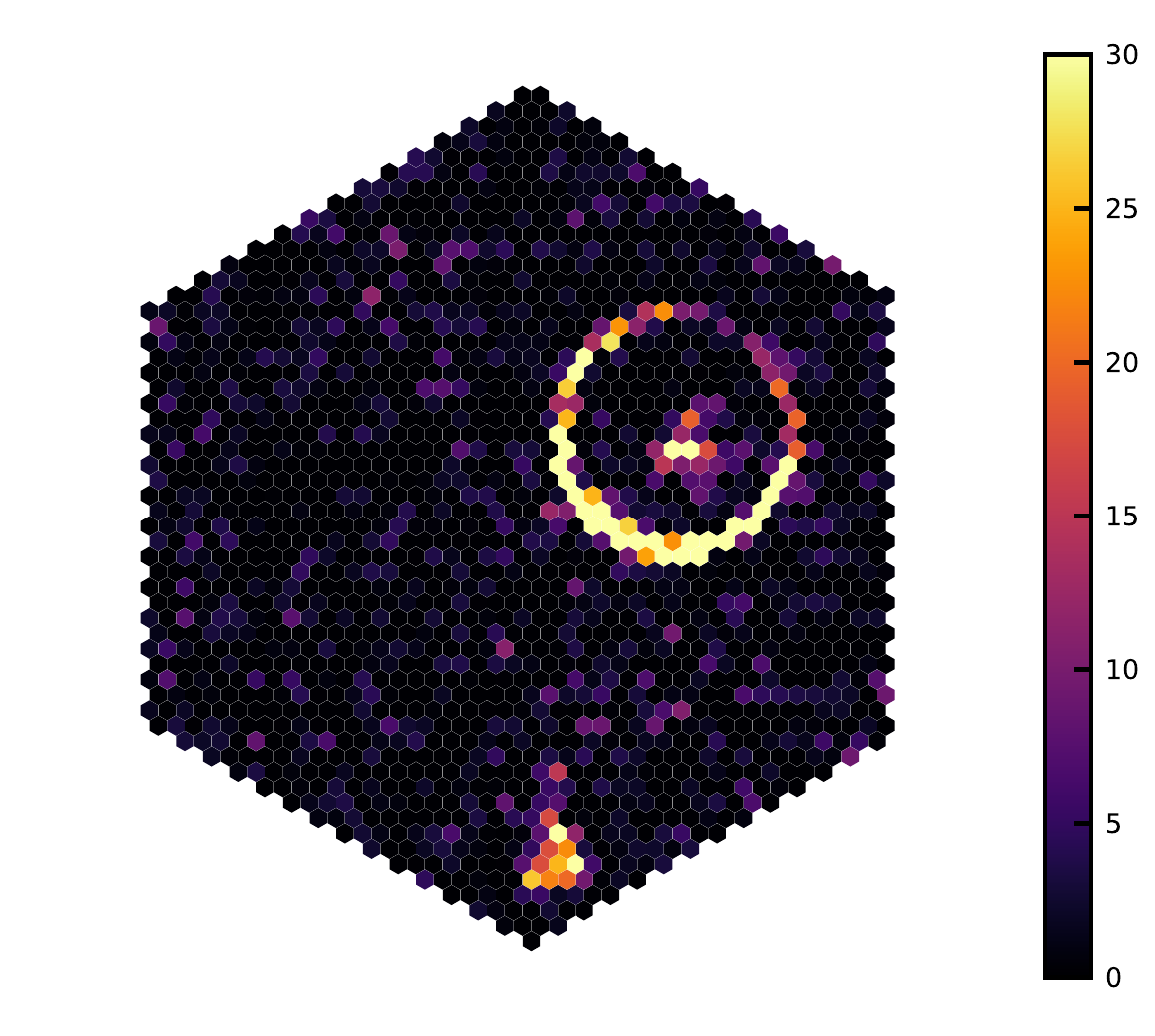}%
\caption{Integrated pixel charge in PEs. }
\label{cw_10}
\end{subfigure}%
\hfill
\begin{subfigure}{\linewidth}
\includegraphics[width=\linewidth]{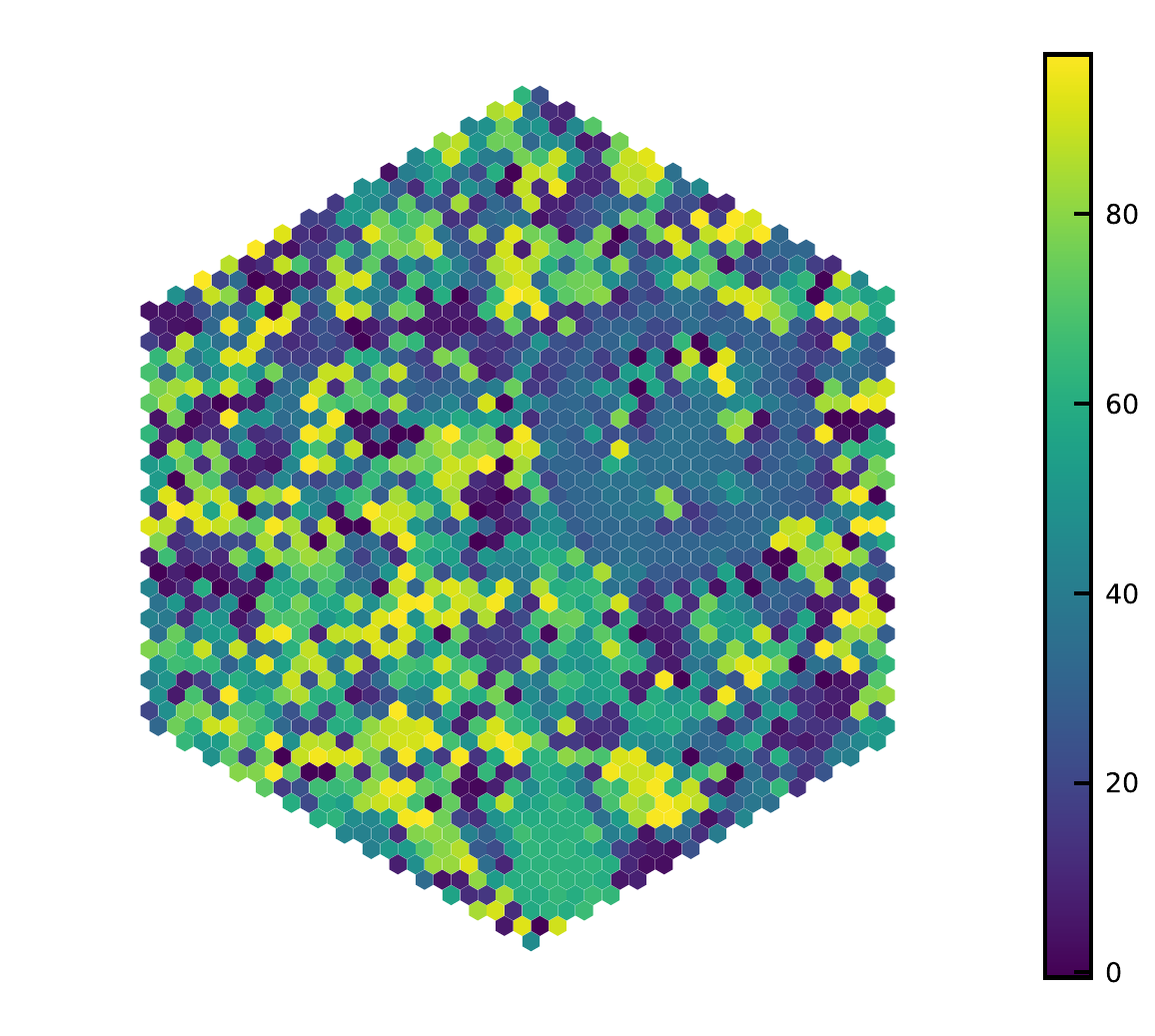}%
\subcaption{Derived pixel arrival time in ns.}
\label{cw_25}
\end{subfigure}
\caption{Example of a simulated proton shower event reconstructed with a neighbour-based extractor. It shows the derived charge~(a) and time~(b) for each pixel. A muon ring can be seen together with a small shower. }
\label{fig:sim_d}
\end{figure}

\subsection{Tailcuts}

Tailcuts~\citep{tailcuts} stands out as one of the foremost {\it Image Cleaning} algorithms widely employed as the default choice across various Cherenkov detectors. Tailcuts method cleans an image by selecting pixels that pass two charge thresholds~(referred here as picture and boundary thresholds). All pixels with a signal exceeding the picture threshold will be kept, provided they have a minimum number of neighbouring pixels. Additionally, all nearest neighbours of a picture pixel that surpass the boundary threshold will also be included~\citep{ctapipe}. There are, therefore, three free parameters: picture and boundary thresholds, and number of picture neighbours. Tailcuts is characterised by its rapid execution, straightforward methodology, and good acceptance of high-significance signal pixels. However, a drawback of Tailcuts is its tendency to discard low-significance pixels, thereby losing crucial information of the shower. Typically situated in the tail of the shower image, these low-charge pixels represent a key aspect of the event. While this is not very problematic for Hillas-based reconstruction methods as they typically operate under the assumption of a shower resembling an ellipse, it can nonetheless result in sub-optimal outcomes for DVR. Consequently, Tailcuts may not align with the CTAO expectations on reduction algorithms, which prioritise the detection of signal pixels and must not affect the reconstruction of events using more complex methods such as ImPACT~\citep{impact}.

Fig.~\ref{fig:sim_d} shows the charge and time distributions of a simulated proton-initiated shower event. The charge distribution clearly shows the distinction between signal and noise, which becomes less clear at the outer layers of the shower. Nevertheless, the time distribution of signal pixels is also very distinct compared to noise. Signal arrives at a similar time in each pixel but noise pixels' trigger times are randomly distributed and uncorrelated to the neighbours. There are, however, some small correlations between nearest neighbours of noise pixels as an effect of the signal extraction algorithm. Tailcuts only makes use of charge information but, in this paper, we study how time could be used to improve pixel selection.

\subsection{Time-based Clustering}

DBSCAN, Density-Based Spatial Clustering of Applications with Noise,~\citep{dbscan} is a density-based clustering algorithm. Clusters are regions in an N-dimensional space of high density separated by points of lower density. DBSCAN is a very well-proven algorithm with two main aspects which are good for {\it Image Cleaning}. Firstly, DBSCAN does not require the user to specify the number of clusters, handling any arbitrary cluster shape. Secondly, points which are not assigned to any cluster are labelled as noise. On the other hand, DBSCAN is also sensitive to data scales, which means that scaling factors may be needed, increasing the number of variables for optimisation. The basic parameters involved in the algorithm are the minimum number of points to consider a dense region (minPts) and the minimum distance to form a neighbourhood (eps). HDBSCAN~\citep{hdbscan}, Hierarchical DBSCAN, extends DBSCAN to a hierarchical clustering algorithm. First tests with HDBSCAN are not promising in terms of performance and signal pixels are generally not well identified. Alternatively, OPTICS~(Ordering Points To Identify the Clustering Structure)~\citep{optics} has also been tested, yielding to similar results with no clear advantage over DBSCAN. Due to its suboptimal computational time, DBSCAN is used throughout the paper.

The amplitude and arrival times have been extracted, as explained earlier, with a neighbour-based signal extractor. Clustering is fed by reconstructed times and positions in the x and y direction of those pixels with charge over a minimum threshold (noise cut)~\citep{time-clustering}. The noise cut is crucial to remove some noise pixels with very low significant pixels which may have time correlations with its neighbour pixels, artificially generated by the signal extractor. Additionally, two scaling factors, spatial and temporal, are applied to the $x$ and $y$ positions and time, respectively. The scales can be chosen so that eps is equal to 1. DBSCAN groups points together based on the proximity of the points. As an output, each point is linked to a cluster ID or noise. 

Furthermore, and to ensure that high significance pixels pass the cleaning, we have included those pixels with a significance higher than an optimised limit and with at least one neighbouring pixel above the same threshold (charge threshold). Lastly, and for any method~(also Tailcuts), rows of pixels may be added around the cleaned image to ensure all the details of the shower are enclosed in the mask. Modifications to this method are studied in \ref{appendix}. None of the methods presented represent the final version of the CTAO DVR algorithm.

\begin{figure}[h!]
    \centering
    \includegraphics[width=\linewidth]{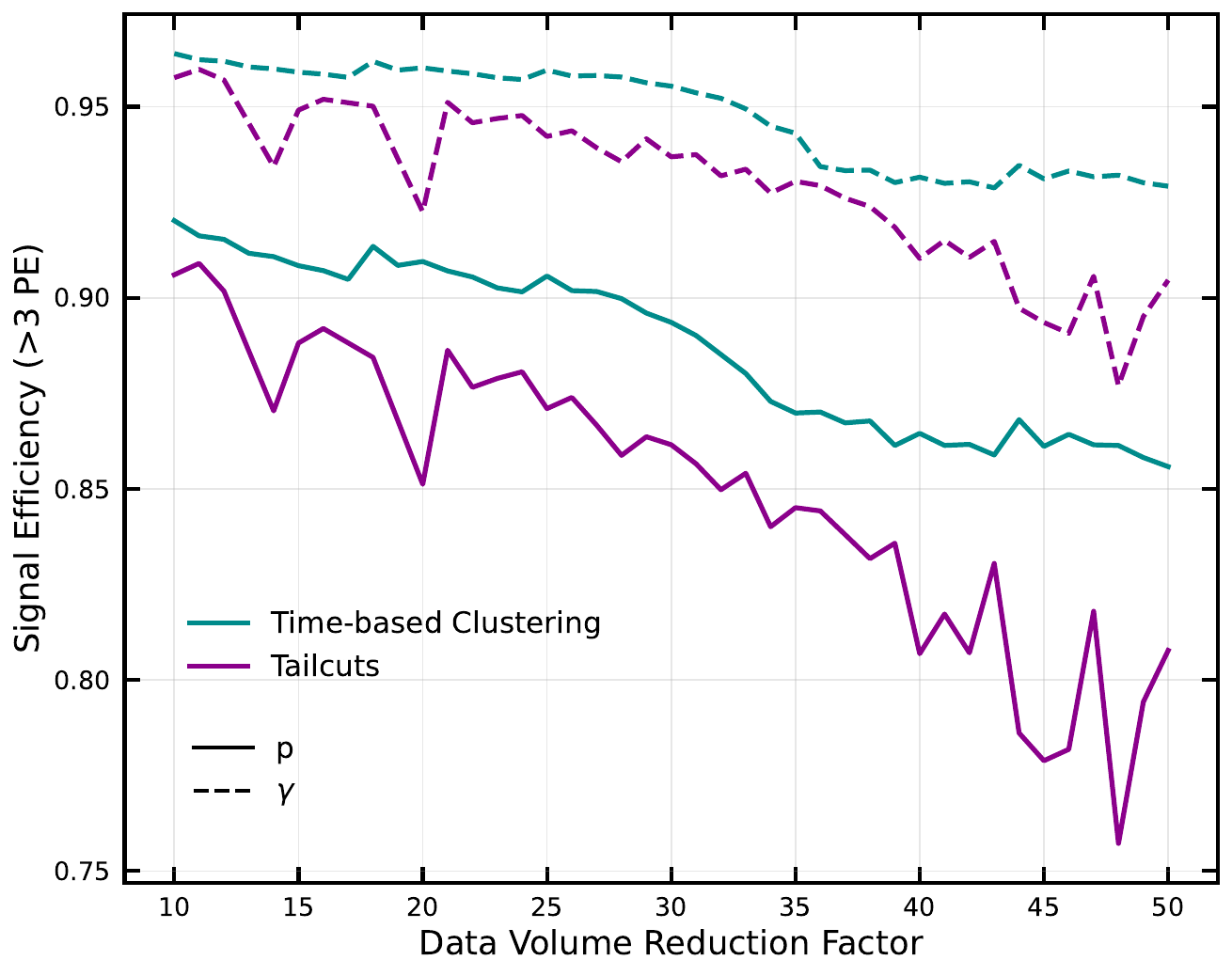}
    \caption{Efficiency for retaining signal pixels (i.e. containing more than three true signal PEs) as the DVR factor increases from 10 to 50 in steps of 1, for both Tailcuts and the Time-based Clustering approach. The signal efficiency is shown for both proton and gamma showers simulations. And the DVR factor, to improve accuracy, is obtained only from proton simulations. } 
    \label{fig:sig_dvr_low}
\end{figure}

\subsection{Waveform cropping}

The second step of DVR, as illustrated in Fig.~\ref{fig:sketch}, is waveform cropping. The majority of events are low-energy and produce short pulses in the camera. The readout window is sufficiently long to fully capture the shower development for long pulses. This results in the storage of waveform samples containing noise for short pulses. However, waveform cropping is not ideal for large, typically long-duration hadronic, showers which produce large pulses that extend over many samples. Reduction of large events has a minimal impact on the DVR factor since high-energy events are less frequent. 

An additional DVR factor of approximately 2 can be achieved by using an event-wise window, fixed for all pixels, to retain only the samples containing relevant information for image extraction. In the following discussion, we will disregard this aspect and focus solely on pixel selection.

\section{Verification}\label{sec:verification}

\begin{figure*}[t!]
     \centering
     \begin{subfigure}[b]{0.3\textwidth}
         \centering
         \includegraphics[scale=0.2]{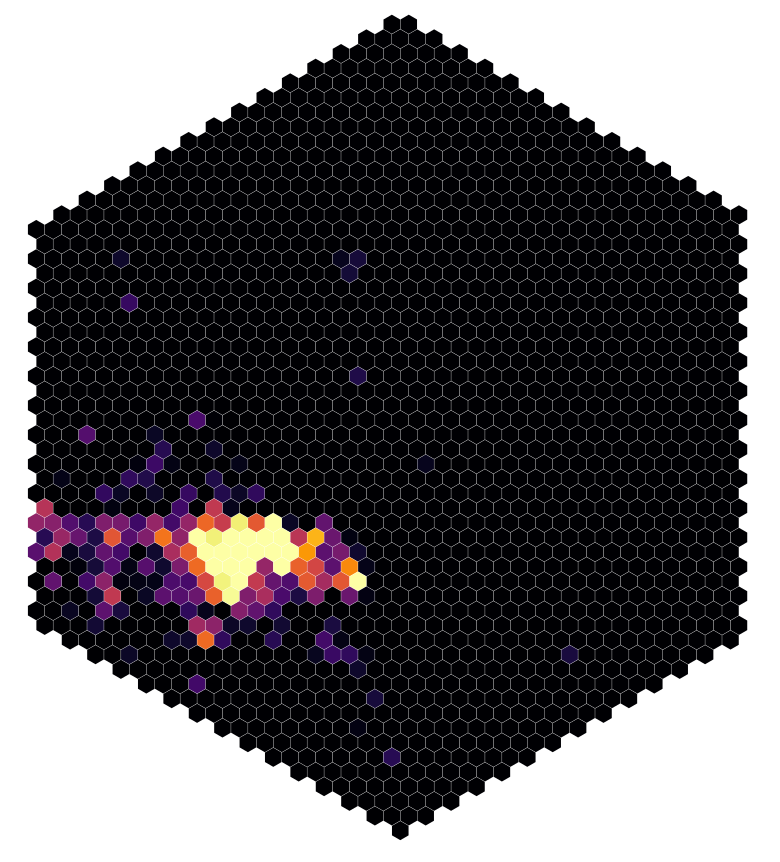}
         \caption{Original}
         \label{original}
     \end{subfigure}
     \hfill
     \begin{subfigure}[b]{0.3\textwidth}
         \centering
         \includegraphics[scale=0.2]{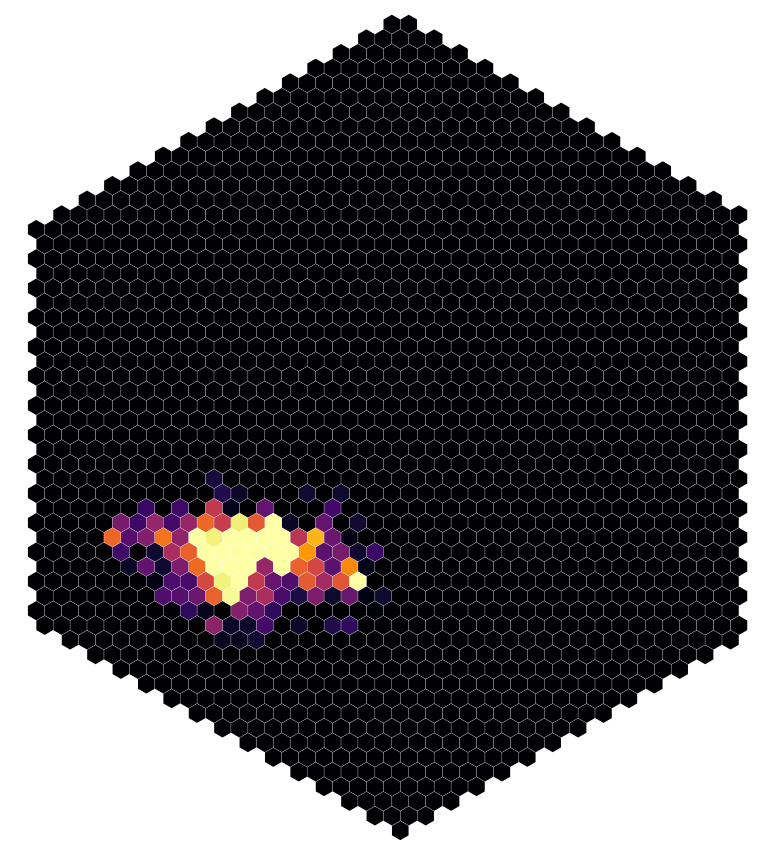}
         \caption{Tailcuts}
         \label{tails}
     \end{subfigure}
     \hfill
     \begin{subfigure}[b]{0.3\textwidth}
         \centering
         \includegraphics[scale=0.2]{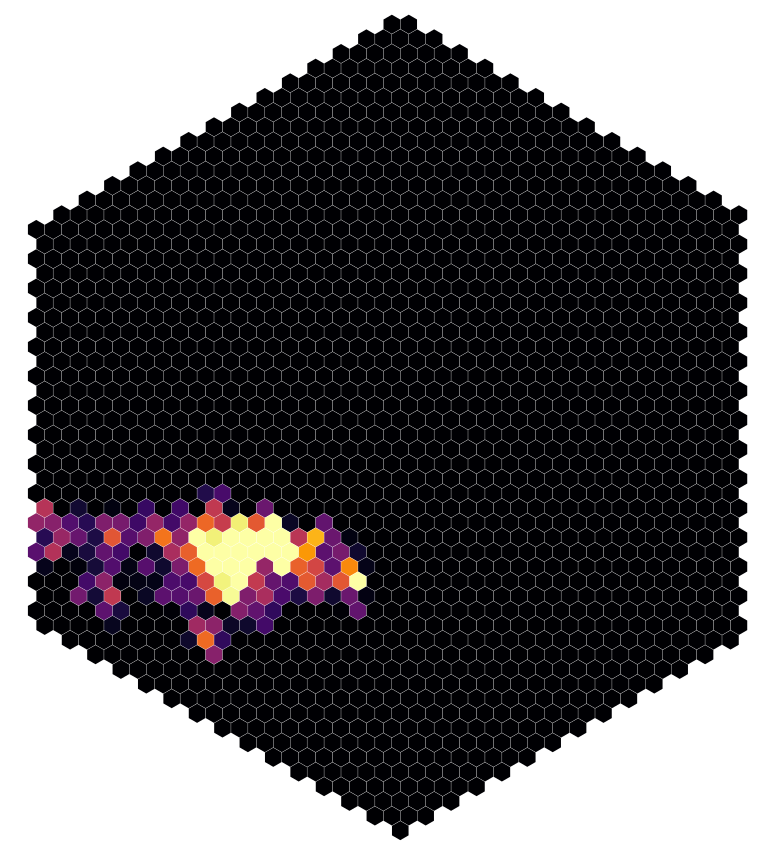}
         \caption{Clustering}
         \label{clust}
     \end{subfigure}
        \caption{Images of a simulated proton shower in an MST-FlashCam before and after DVR. Two DVR algorithms, Tailcuts and Clustering, were employed, both configured to achieve a DVR factor of 40. The colour scale of the plots ranges from 0 to 15\,PEs. The purple region of the shower tail represents values around 3\,PEs. }
        \label{fig:distribs}
\end{figure*}

In this section we compare the performance of the two algorithms introduced above~(Tailcuts and Time-based Clustering), firstly at the level of individual camera images, and secondly at the level of the array, i.e. on gamma-ray reconstruction and background rejection. For the verification and comparison of algorithms we focus on three main points:
\begin{itemize}
    \item the fraction of true signal pixels that are identified as such at a given reduction factor;
    \item scientific performance achieved at high reduction factors, when applying the same algorithm also for Imaging Cleaning;
    \item response of the algorithm to increasing NSB, number of broken pixels, and calibration uncertainty.
\end{itemize}

These tests are performed with simulations of cosmic and gamma ray showers. Showers are generated with {\it CORSIKA}~\citep{corsika} and the telescope array is simulated with {\it sim\_telarray}~\citep{simtelarray}. Simulations are made with a preliminary version of Prod-6, the latest available candidate configuration for the next large-scale simulation production. We simulate the South site of the CTAO with 14 MSTs~\citep{alpha-config} with the layout showed in \ref{appendix_layout} at 20\(^\circ\) zenith angle. Events have been simulated with an energy spectrum of $\frac{dN}{dE} \propto E^{-2}$. All events have been weighted by $E^{-0.7}$ to achieve an energy distribution $\propto E^{-2.7}$. Events are simulated with a nominal NSB rate of 0.216\,PE/ns per pixel.

\subsection{Image-level performance}

The basic performance of a DVR algorithm can be evaluated by examining the fraction of true signal pixels correctly identified at a given data reduction factor. In this study, we analyse how signal efficiency evolves with increasing data reduction using simulations of around 10,000 proton- and 30,000 gamma-initiated showers with energies between 0.01 and 310\,TeV.

Fig.~\ref{fig:sig_dvr_low} illustrates the signal efficiency for pixels with more than 3\,PEs as a function of the DVR factor, comparing the two previously introduced methods. The DVR factor is defined as the number of pixels before reduction divided by the number of pixels selected after DVR. Each data point in the plot corresponds to a unique set of free parameters. For each DVR factor, more than one set of free parameters are typically seen but only the point with higher signal efficiency is selected and plotted for each method. 

The clustering algorithm consists of 6 free parameters for optimisation. These include spatial and temporal scales, minPts, noise cut, charge threshold, and the number of dilated rows~(ranging from 0 to 2) added around the image after pixel reduction. For Tailcuts, the free parameters are the high (picture) and low (boundary) thresholds, the minimum number of picture neighbours, and the number of dilated rows. 

Clustering consistently detects more signal than Tailcuts across the entire range of data reduction values for both gamma and proton showers. Specifically, Clustering always identifies more than 85\% of signal pixels above 3\,PEs for proton showers, while Tailcuts drops to 75\%. Images created by gamma-ray showers are generally more compact and less scattered~\citep{aharonian1997potshowers} than those from proton showers, making the signal pixels from gamma showers easier to detect and resulting in higher signal efficiencies of around 95\% for Clustering.

The optimal parameters at a DVR factor of 40 were selected from Fig.~\ref{fig:sig_dvr_low} to generate the images shown in Fig.~\ref{fig:distribs}. The shower depicted in Fig.~\ref{clust}~(Clustering) exhibits more features and retains more low-charge pixels compared to the one in Fig.~\ref{tails}~(Tailcuts), which preserves most of the shower but loses crucial information. Table~\ref{tab:clust_parameters}~and~\ref{tab:tails_parameters} provide the parameters selected for the two DVR algorithms used, at the four DVR levels adopted in this study.

\begin{table}
    \centering
    \begin{tabular}{|c|c|c|c|c|} \hline
     \textbf{DVR factor} & {\bf 30} & {\bf 40} & {\bf 120} & {\bf 140} \\
     \hline
     Spatial scale\,(m) & 0.175 & 0.175 & 0.175 & 0.15 \\
     \hline
     Time scale\,(ns) & 5.0 & 4.0 & 5.0 & 5.0 \\
     \hline
     minPts & 6 & 4 & 5 & 7 \\
     \hline
     Noise cut\,(PE) & 2.3 & 2.5 & 2.3 & 2.3 \\
     \hline
     Charge threshold\,(PE) & 8.0 & 6.0 & 10.0 & 8.0 \\
     \hline
     Number rows & 2 & 1 & 0 & 0 \\
     \hline
    \end{tabular}
    \caption{The Time-based Clustering algorithm configurations for MST-FlashCam are optimized at four DVR factors: 30, 40, 120, and 140. These parameters, selected based on Fig.~\ref{fig:sig_dvr_low}, are chosen to maximize the signal efficiency of pixels with signals exceeding 3\,PEs at each DVR factor. }
    \label{tab:clust_parameters}
\end{table}

\begin{table}
    \centering
    \begin{tabular}{|c|c|c|c|c|} \hline
     \textbf{DVR factor} & {\bf 30} & {\bf 40} & {\bf 120} & {\bf 140} \\
     \hline
     Picture threshold\,(PE) & 4.0 & 6.0 & 5.0 & 12.0 \\
     \hline
     Boundary threshold\,(PE) & 5.0 & 4.0 & 3.0 & 2.0 \\
     \hline
     Picture neighbours & 2 & 2 & 1 & 0 \\
     \hline
     Number rows & 2 & 2 & 0 & 0 \\
     \hline
    \end{tabular}
    \caption{Configurations of Tailcuts algorithm at the main four DVR factors used in this paper for MST-FlashCam: 30, 40, 120, and 140. }
    \label{tab:tails_parameters}
\end{table}

\subsection{Array-level performance}

Reconstruction performance plots are generated with simulations, previously described in detail, of $10^8$ simulated proton and gamma showers within a maximum radius of 1000\,metres. From shower simulations, charge and time information are extracted, followed by cleaning using either Tailcuts or Time-based Clustering methods, configured with DVR factors of 120 and 140. Effective {\it Image Cleaning} demands higher noise rejection to achieve accurate Hillas-based signal reconstruction, which requires increased DVR factors. We selected these two DVR values to demonstrate how reconstruction resolution varies when similar data reduction is achieved.

\begin{figure}[h!]
\begin{subfigure}{\linewidth}
\includegraphics[width=\linewidth]{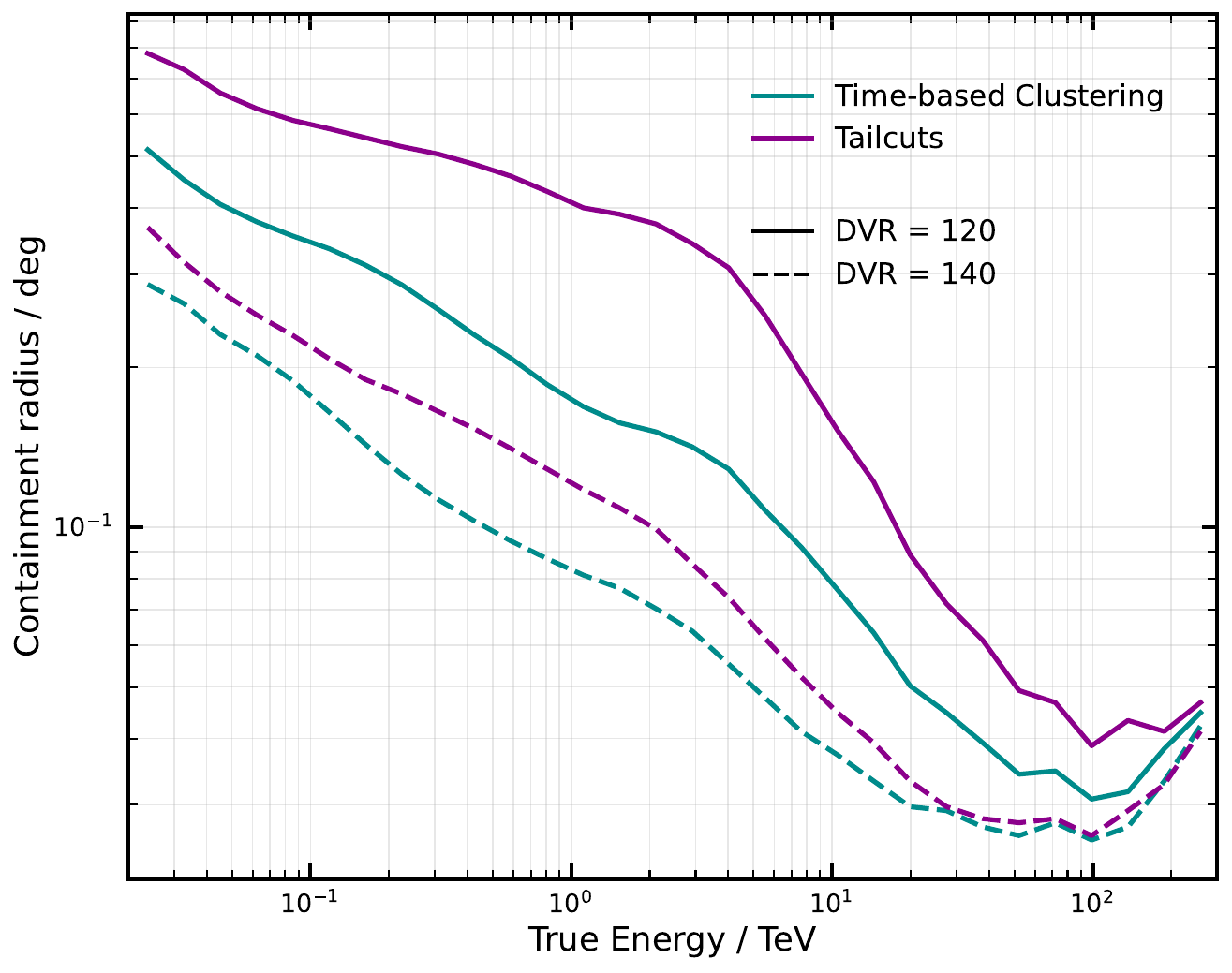}%
\caption{Angular resolution. }
\label{angular}
\end{subfigure}%
\hfill
\begin{subfigure}{\linewidth}
\includegraphics[width=\linewidth]{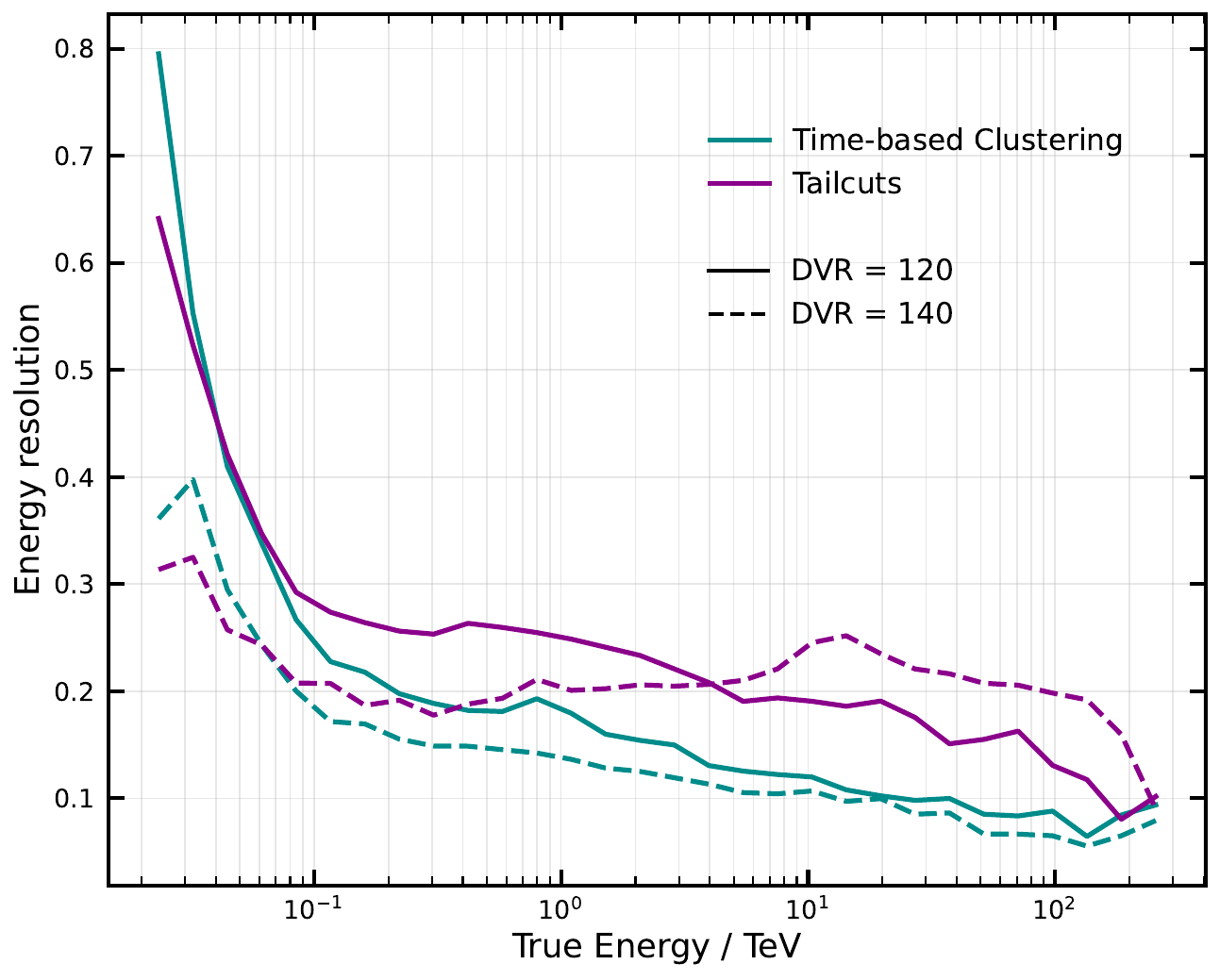}%
\subcaption{Energy resolution.}
\label{energy}
\end{subfigure}
\caption{Angular and energy resolution is shown as a function of the true gamma-ray energy. The two methods are compared using DVR factors of 120 and 140. Higher DVR factors are selected to improve reconstruction quality. These values are chosen to be close to each other, allowing for a comparison of how the resolution evolves with a small increase in data reduction. }
\label{fig:perf}
\end{figure}

After the cleaning routines, the Hillas parameters are extracted in a consistent manner. The shower reconstruction is performed with a Random Forest. The reconstructed direction is obtained with the geometrical Hillas method~\citep{ang_rec}. The energy is reconstructed by training a Random Forest regression implemented in {\it ctapipe}~\citep{ctapipe}. We train the regression algorithm with 14 parameters: Hillas parameters (width, length, area, and intensity), fraction of total charge after cleaning concentrated in the centre of gravity of the shower image, fraction of charge in core, number of clusters, impact distance, maximum height, percentage of shower intensity on the outer layers of the camera, maximum intensity, mean intensity and standard deviation of the intensity, and number of pixels after cleaning. 

Furthermore, and to ensure the good quality of the fit, only events that trigger at least three telescope after quality cuts are used for reconstruction. The quality cuts require camera images to pass the following limits. First, at least three pixels should survive the cleaning. Second, the Hillas intensity~(total charge after cleaning) must be greater than 50\,PEs. The two first cuts assure that images are sufficient bright for accurate reconstruction. Third, events must have valid Hillas parameters and non-zero Hillas width for reconstruction to be valid. Last, more than 20\% of the shower intensity cannot be located at the outer two rows of pixels of the camera. This last cut is applied to limit image truncation and include only those images which are developed (almost) completely inside the physical boundaries of the camera. Reconstruction performance could, otherwise, be degraded.

Results from the Hillas-based reconstruction serve as seeds for other methods like ImPACT. To improve the performance of ImPACT, the resulting cleaned image is dilated by a few rows of pixels, typically 2, that will be used for the training of the data and the reconstruction. The key information for ImPACT is in the tails of the image. Hillas-based reconstruction, on the other hand, benefits from higher pixel reductions and ellipse-like images. After DVR, images need to conserve enough pixels and information so that methods like ImPACT do not see a drop in their performance due to an excessively large pixel reduction. 

Fig.~\ref{angular} shows the angular resolution of gamma showers as a function of true energy. The angular resolution is defined as 68\% containment of the distribution of the angular distance from the source position~\citep{gammapy}. A clear improvement of Clustering over Tailcuts is apparent over the entire energy range at both DVR factors: 120 and 140. We see a significant degradation in performance at 120 compared to 140. This is a clear example of how the increase in data reduction helps event reconstruction when Hillas-based methods are used. The free parameters of the cleaning method are selected to maximize signal efficiency for a given reduction factor. However, this approach does not necessarily lead to the optimal parameter choice for achieving the best resolution. Ideally, each parameter set should be tested throughout the entire analysis process to identify the parameter combination that optimizes resolution.

Fig.~\ref{energy} shows the energy resolution as a function of the true energy of the simulated gamma showers. Energy resolution is defined at half-width of the 68\% interval around 0 of \(\frac{\text{E}_{\text{reco}} - \text{E}_{\text{true}}}{\text{E}_{\text{true}}}\), where \(\text{E}_{\text{reco}}\) and \(\text{E}_{\text{true}}\) refers to the reconstructed and simulated energy of the initiated gamma particle, respectively. A modest improvement of Clustering over Tailcuts is seen over almost the whole energy range at both DVR factors. At low energy levels, the energy reconstruction is inaccurate, resulting in unreliable outcomes.

\subsection{Computation time}

The current implementation of the algorithms is done using Python~\citep{python}, and as of now, their performance is not optimised for speed. However, initial performance checks show that the Clustering and Tailcuts methods exhibit similar computation times. Both approaches are able to process between 200 and 300 events per second when running on a single CPU core, which would need up to about 20 cores to reach the CTAO requirements. Although these results are promising for a basic implementation, there is significant potential for improvement.

\section{Robustness} \label{sec:robustness}

\subsection{Night Sky Background Light}

Reconstructing gamma-ray signals in the presence of high NSB is challenging. However, observing the sky under conditions of high or inhomogeneous NSB, and during moonlight, is quite common and must be addressed appropriately. To assess the impact of NSB on reconstruction performance, we simulate a FlashCam camera exposed exclusively to NSB, without any laser or Cherenkov light illumination. These simulations are conducted using {\it sim\_telarray}, configured to trigger on events in the absence of air showers. 

\begin{figure}
    \centering
    \includegraphics[width=\linewidth]{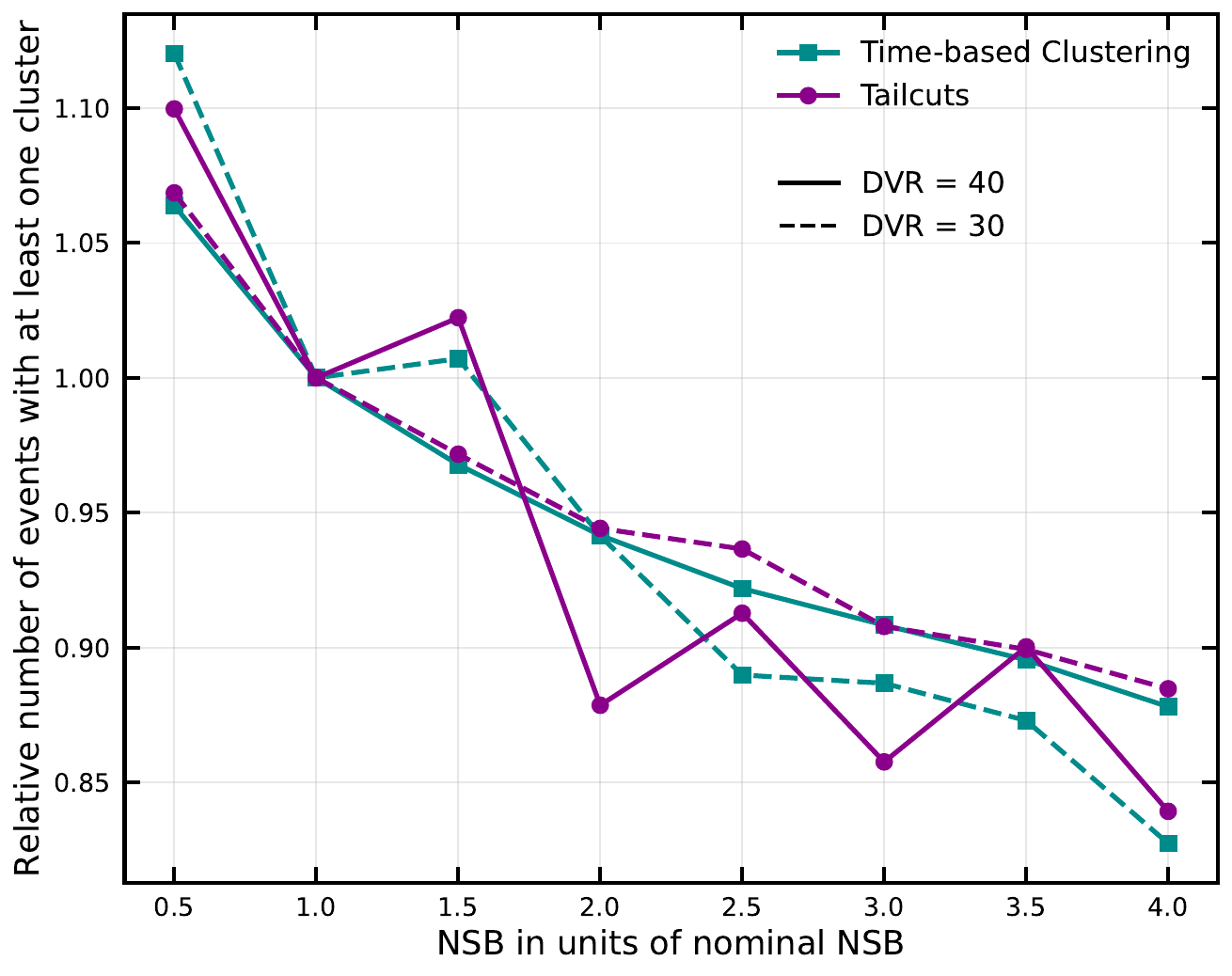}
    \caption{The ratio of the number of NSB-only events with at least one misidentified noise cluster to the number of such events at nominal NSB~(0.216 PE/ns) is examined as a function of increasing NSB rate. The two DVR algorithms are evaluated at DVR factors of 30 and 40.}
    \label{fig:noise_clusters}
\end{figure}

The charge-dependent thresholds for both methods in this section are defined not in terms of absolute charge in PE, as presented in Table~\ref{tab:clust_parameters}, but rather as quantiles of the charge distribution for each pixel across all events. The charge distribution is obtained for each pixel individually to account for inhomogeneities and bright stars. The charge thresholds are obtained by calculating the quantile, fixed for every NSB level, of each distribution. Consequently, the charge-dependent thresholds for both methods, Tailcuts and Clustering, increase with NSB.

Fig.~\ref{fig:noise_clusters} illustrates how the number of events with at least one noise cluster after cleaning varies as the NSB photon rate increases. All lines in the plot are normalised to the number of events with noise clusters at the nominal NSB rate of 0.216\,PE/ns for FlashCam. The methods demonstrate a modest dependency on the NSB level, with a similar trend observed for both methods. The rate of events with noise decreases because in the neighbour-based charge reconstruction the probability to find a pair of unusually large noise pulses (e.g., exceeding the 99th percentile of the individual pixel charge distribution) in neighbouring pixels decreases with NSB. This becomes visible in both Tailcuts and Time-based Clustering finding fewer clusters.

At a DVR factor of 30, both methods have the same number of rows added around the cleaned image (Table~\ref{tab:clust_parameters}~and~\ref{tab:tails_parameters}). Clustering, however, identifies only 3\% of the events, while Tailcuts detects roughly an order of magnitude more. At a threshold of 40, with 2 dilated rows for Tailcuts and 1 for Clustering, Tailcuts misidentifies clusters in just 1\% of the events, whereas Clustering incorrectly classifies clusters in 40\%. Fig. \ref{fig:sig_dvr_low} only shows the best parameter choice for each DVR factor based on signal efficiency. Nevertheless, the choice of free parameters could also be obtained based on the number of noise clusters if that is beneficial for a specific study. The number of noise clusters can be reduced, for example, by increasing the number of rows, minPts, or the picture threshold and neighbours.

\subsection{Broken Pixels}
\label{sec:broken}

Cherenkov cameras may experience pixel deactivation due to hardware malfunctions or very bright stars. These so-called broken (or unavailable) pixels return no signal. Bright stars commonly manifest as one or two deactivated pixels, for safety precautions, accompanied by one or two rows of bright pixels exhibiting elevated NSB levels compared to the average. H.E.S.S. has studies on the effect of bright stars~\citep{bright-stars}. 

\begin{figure}[h!]
    \centering
    \includegraphics[width=\linewidth]{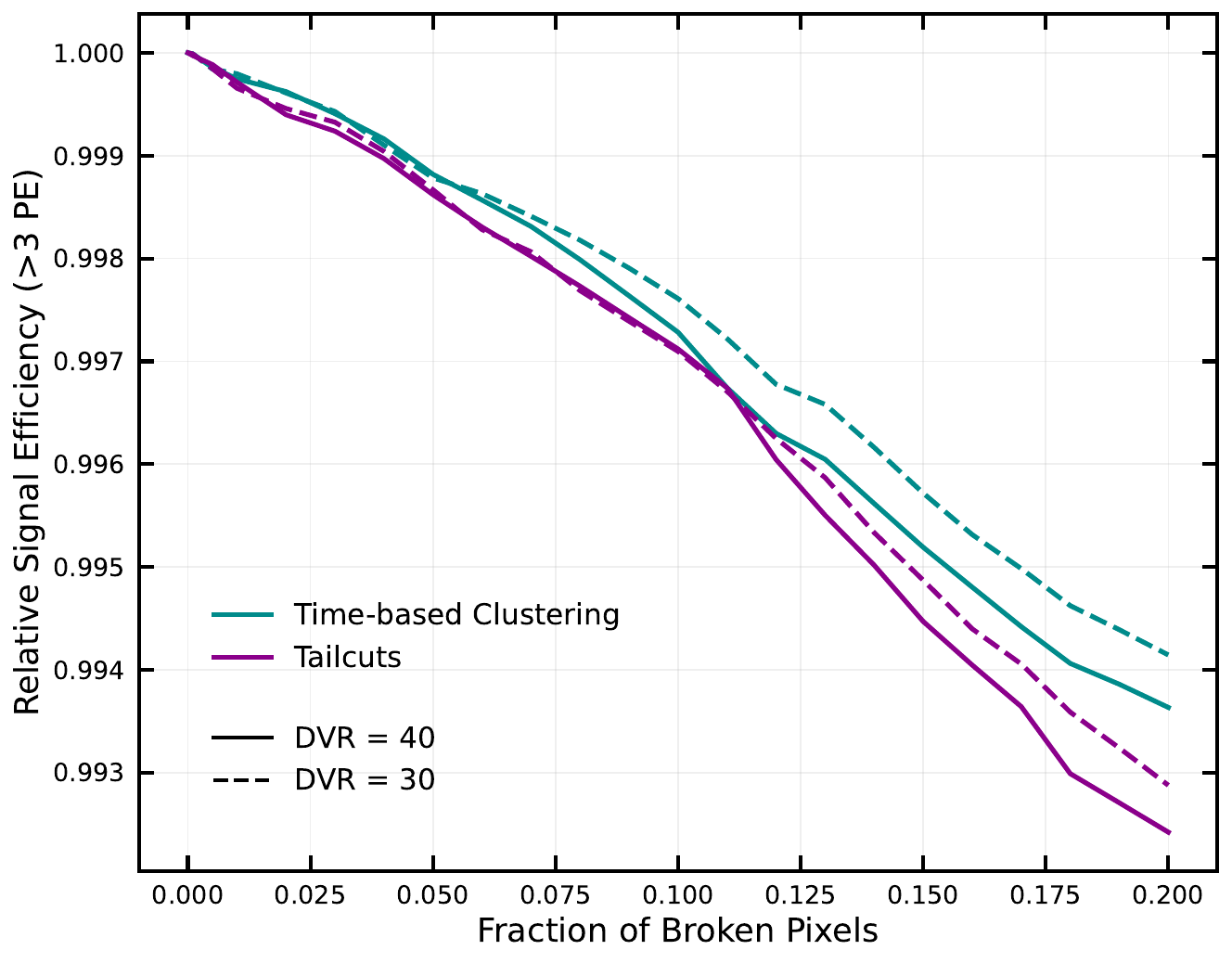}
    \caption{Fraction of signal pixels detected with more than 3\,PEs with increasing number of broken pixels for simulated gamma rays divided by the efficiency calculated from the standard, no broken pixels, simulation (which is different for the two algorithms, cf.~Fig.~\ref{fig:sig_dvr_low}). Two methods are presented: Tailcuts and Clustering configured for DVR factors of 30 and 40. }
    \label{fig:intensity_bps}
\end{figure}

\begin{figure}[h!]
    \centering
    \includegraphics[width=\linewidth]{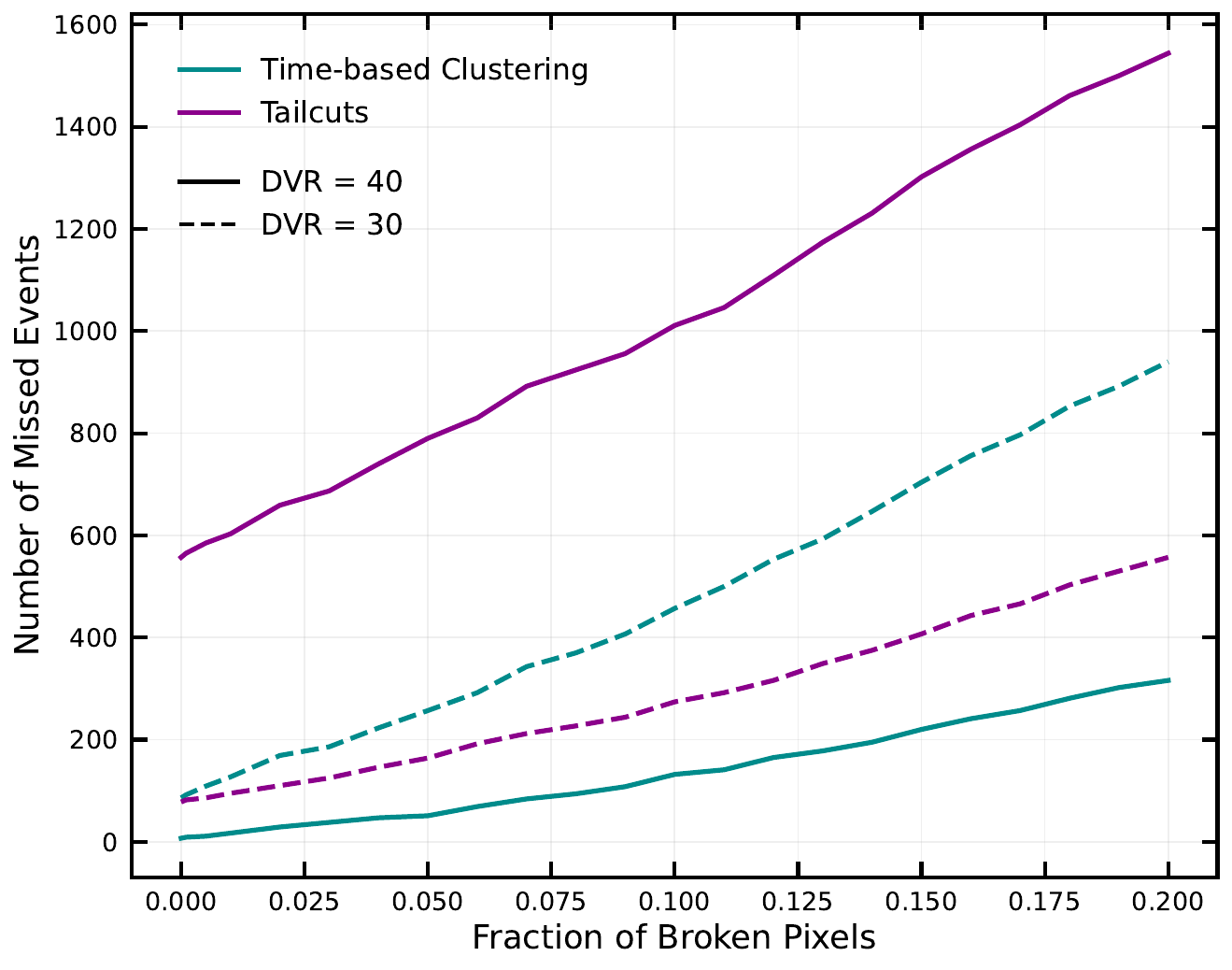}
    \caption{Number of events missed after DVR with Tailcuts and Clustering at 30 and 40, as the number of broken pixels increases. }
    \label{fig:missed_events}
\end{figure}

Deactivated pixels are simulated with zero intensity. Pixel-wise algorithms cannot simply ignored these pixels as they influence the pixel selection. Therefore, the zero intensity and time of broken pixels is substituted by the average charge and trigger time of its neighbouring pixels. This way, pixels of zero intensity, which could heavily influence the performance of the method, are filled by the average of the nearest neighbours. Fig.~\ref{fig:intensity_bps} shows the loss in signal efficiency, for pixels with more than 3\,PEs, when a fraction of broken pixels randomly located in the camera are deactivated. The signal efficiency in the y-axis is divided by the intensity of the same exact events but no broken pixels to improve the visibility of the plot. Clustering always detects a higher number of signal pixels as shown in~Fig.~\ref{fig:sig_dvr_low}. 

We analyse the intensity loss incurred by both Tailcuts and Clustering algorithms across two different DVR factors. Tailcuts and Clustering keep more than 97\% of the signal pixels detected at normal conditions, even at 20\% (more than 300 dead pixels). Clustering performs slightly better than Tailcuts with smaller deviations from the signal efficiency at a fraction of broken pixels of 0. Therefore, the change in signal efficiency amounts to only a marginal percentage for both methods and different DVR factors, indicating negligible impact.

Fig.~\ref{fig:missed_events} illustrates the relationship between the number of missed events and the fraction of broken pixels. Note that more than 20,000~events trigger the camera. The Tailcuts algorithm shows a much higher number of missed events at a DVR factor of 40 than at 30, which is attributed to a higher picture threshold~(See Table~\ref{tab:tails_parameters}). This shows that a small change in the picture threshold can lead to big changes in terms of missed events. In contrast, the Clustering algorithm at a DVR factor of 40 successfully detects nearly all events when operating under the standard configuration~(0.0 fraction of broken pixels). However, at a DVR factor of 30, the Clustering algorithm misses more events due to increased minPts and charge thresholds, Table~\ref{tab:clust_parameters}. The number of missed events with Tailcuts shows less variation compared to Clustering at a DVR factor of 30. The trigger settings are identical for each line. Adjusting these settings could lead to a change in the number of missed events. Overall, simulated broken pixels do not degrade the performance of the algorithms significantly. 

\subsection{Calibration Uncertainty}

We investigate the impact of adding gain calibration uncertainty of up to 50\% on the two presented methods compared to the performance of these methods when no uncertainty is added to the simulations. This uncertainty is simulated by multiplying each pixel waveform by a Gaussian random number with zero mean and the desired standard deviation.

\begin{figure}[h!]
    \centering
    \includegraphics[width=\linewidth]{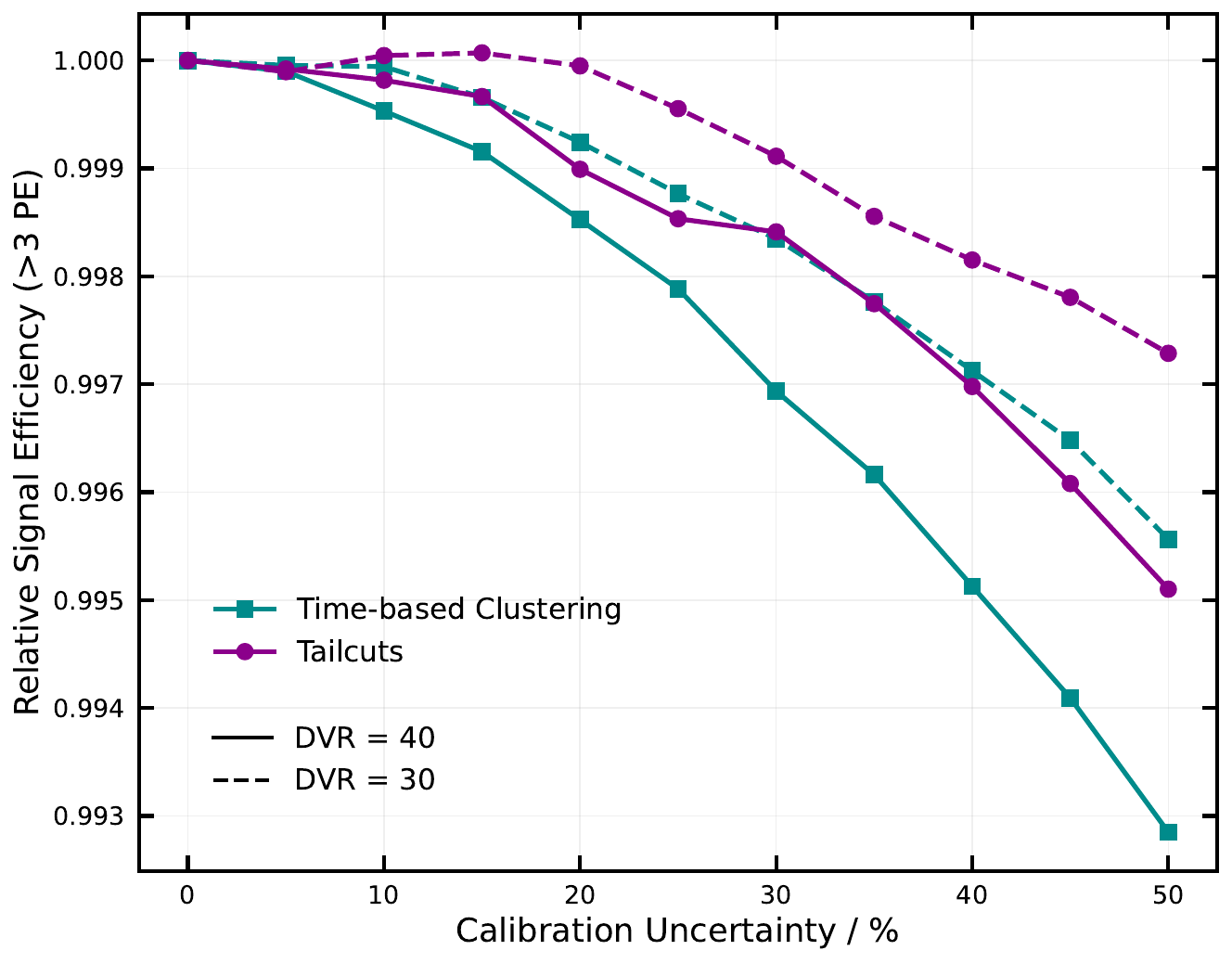}
    \caption{Fraction of signal pixels detected with more than 3\,PEs with increasing calibration uncertainty for simulated gamma rays divided by the efficiency calculated from the standard, no added miscalibration, simulation. Two methods are presented: Tailcuts and Clustering configured for DVR factors of 30 and 40. }
    \label{fig:calib_uncert}
\end{figure}

The y-axis of Fig.~\ref{fig:calib_uncert} shows the efficiency of detecting signal pixels with more than 3\,PEs divided by the efficiency when no calibration uncertainty is added, the standard {\it sim\_telarray} configuration. 

Fig.~\ref{fig:calib_uncert} compares the effects of calibration uncertainty on events cleaned by the two different methods. These methods are evaluated at two reduction factors. At a reduction factor of 30, both methods exhibit similarly good performance, with Tailcuts showing a slightly better performance than Clustering. At a reduction factor of 40, we observe larger shifts in both methods, although the performance degradation is still very small. It is important to point out that Clustering always detects more signal pixels.

Overall, both methods demonstrate a very high level of robustness against calibration uncertainty and broken pixels. The effect is negligible in all the tested cases. The results are promising but robustness should be fully assessed using observational data to validate these findings.

\section{Other cameras} \label{sec:cameras}

The CTAO will consist of three different telescope types with four different cameras. Besides MST-FlashCam, which was investigated in detail above, the method must be applicable to the other cameras as well: the SST camera~\citep{sst}, MST-NectarCAM~\citep{nectarcam}, and the LST camera~\citep{lst}. For all cameras, signals are extracted using a neighbour-based image extractor which integrates the waveform around a peak found by the average of the nearest neighbours. This is similar to the extraction method of FlashCam but without waveform interpolation or deconvolution, which are not needed due to the faster sampling rate of these cameras.

\begin{figure}[h!]
    \centering
    \includegraphics[width=\linewidth]{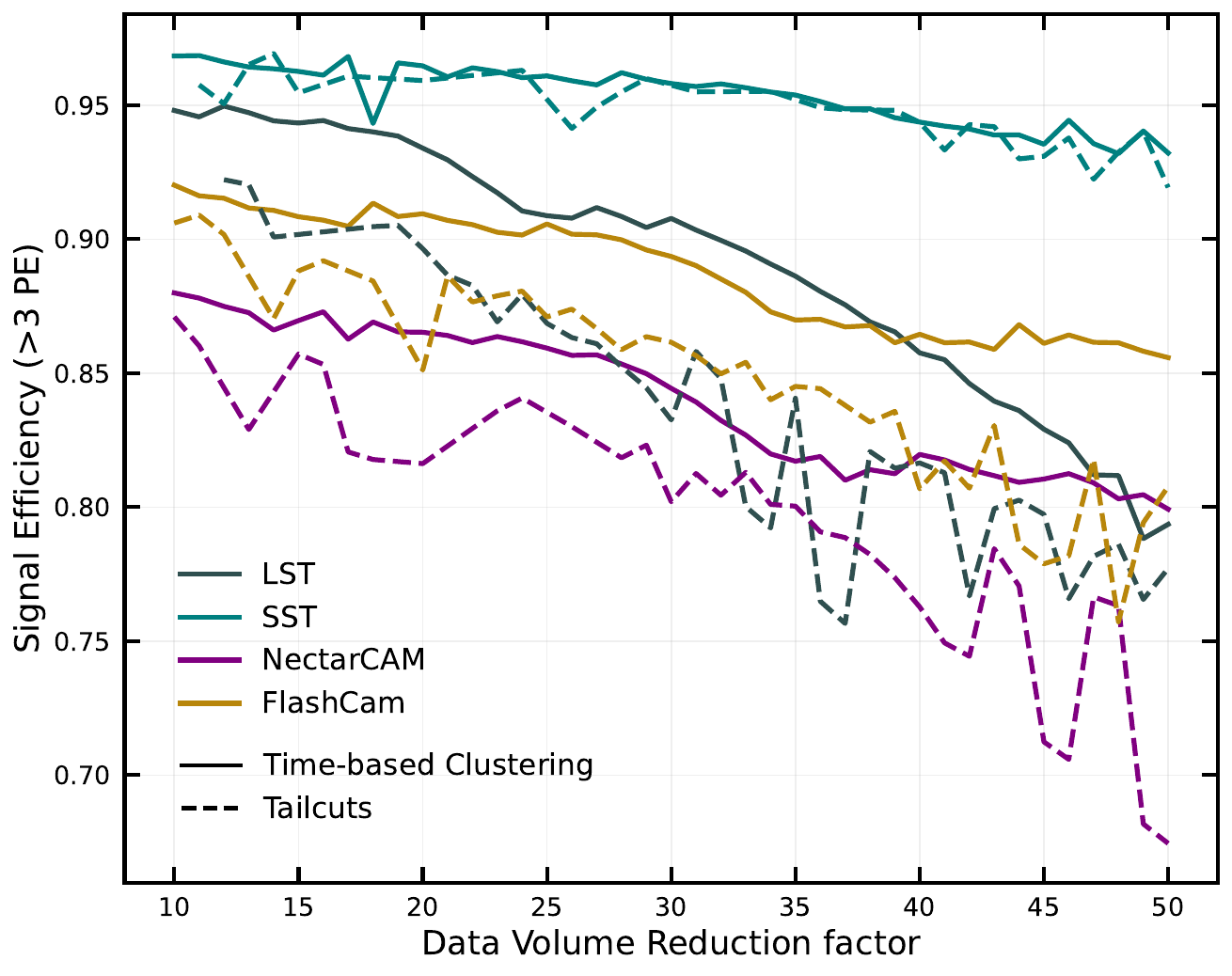}
    \caption{Detection efficiency of signal pixels with more than 3\,PEs as the DVR factor increases from 10 to 50 for the four Cherenkov cameras: SST-Camera~(NSB 0.027\,PE/ns), LST-Camera~(0.198\,PE/ns), and MST-FlashCam and MST-NectarCAM~(0.216\,PE/ns). Signal efficiency and DVR factor calculated from proton shower simulations. Dashed lines show the performance of Tailcuts and the solid lines of Clustering. }
    \label{fig:sig_dvr_cams}
\end{figure}

Fig.~\ref{fig:sig_dvr_cams} shows the signal efficiency for detecting pixels with more than 3\,PEs from proton showers with SST-Camera, LST-Camera, NectarCAM, and FlashCam. Both Tailcuts and Clustering exhibit similar behaviour. Clustering performs better at detecting signal pixels at the same reduction factor, i.e.\ always detects more signal and less noise for NectarCAM, FlashCam, and LST-Camera. The SST camera presents a very high signal efficiency using any of the two methods. Note that the quantisation of some parameters in Tailcuts, picture and boundary thresholds, leads to jumps in the signal efficiency as a function of DVR. 

\begin{figure}[h!]
    \centering
    \includegraphics[width=\linewidth]{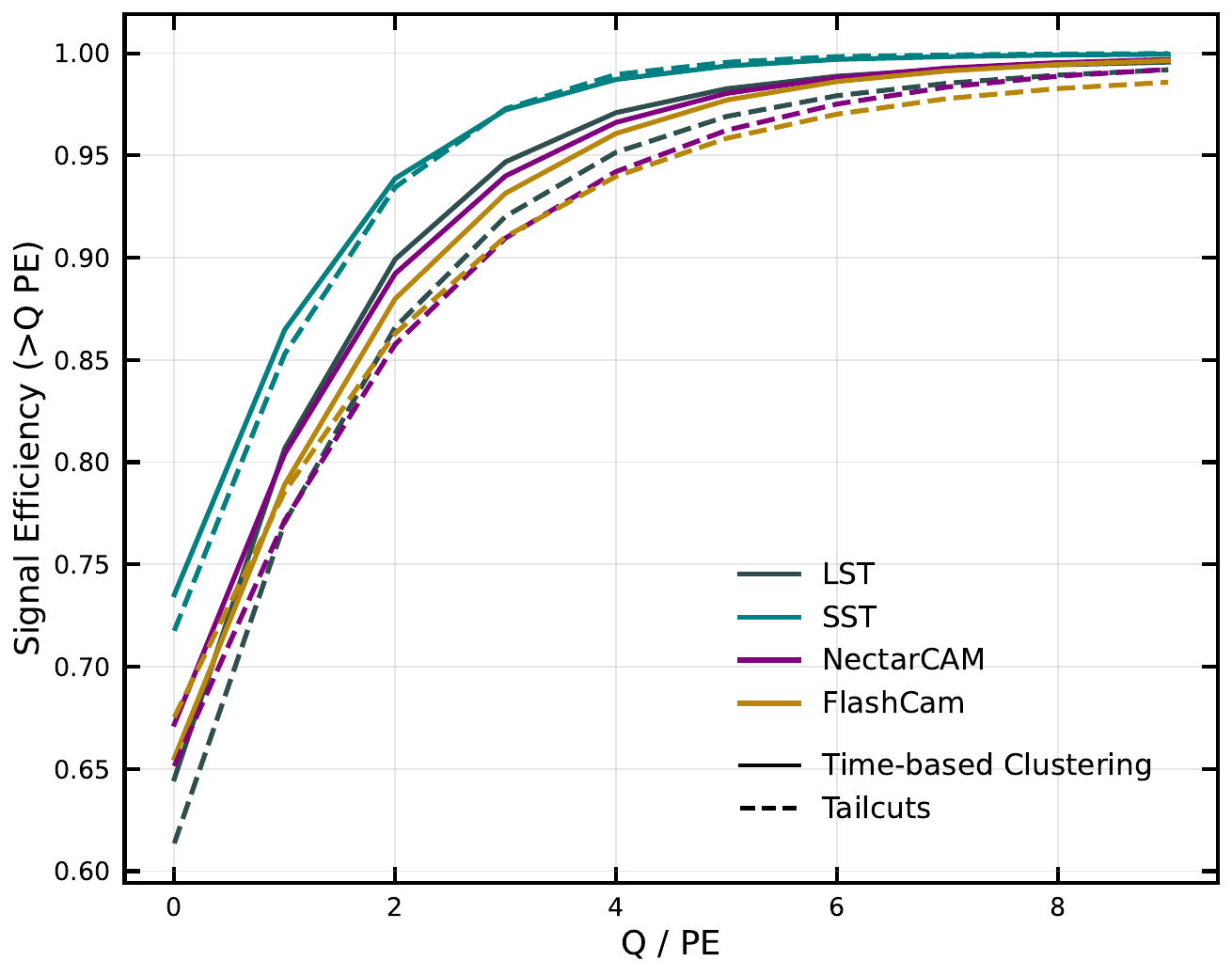}
    \caption{Detection efficiency of signal pixels with more than $Q$\,PEs as a function of $Q$ using gamma showers. The set of parameters for Clustering and Tailcuts are chosen for a DVR factor of 40. Four Cherenkov cameras are explored: SST-Camera, LST-Camera, MST-NectarCAM, and MST-FlashCam. Dashed lines show the performance of Tailcuts compared to the solid lines (Clustering). }
    \label{fig:sig_charge_cams}
\end{figure}

Fig.~\ref{fig:sig_charge_cams} shows the signal efficiency of detecting more than a number of PEs~($Q$) as a function of $Q$ using gamma-initiated showers for all four cameras. The SST camera shows a better signal efficiency than all the other cameras reaching an efficiency very close to 1 at 4\,PEs due to the low nominal level of NSB in each pixel. Clustering for SST is better at low $Q$ and slightly worse than Tailcuts at higher $Q$, but overall the performance is nearly identical. LST, FlashCam, and NectarCAM have similar performance at detecting signal pixels of gamma-ray showers and Clustering overall detects more signal pixels than Tailcuts for all of the three cameras.

\section{Conclusion}
\label{sec:conclusion}

This study indicates that a DBSCAN-based data reduction, following a simple but effective waveform processing step, can provide very robust and effective means of reducing the volume of data stored for the CTAO and indeed for IACT arrays in general. Extensively tested for the FlashCam system, we have shown that this approach is also effective for the very different cameras of the Small- and Large-sized Telescopes, and also for NectarCAM. The Clustering method achieves superior signal efficiency and lower noise detection compared to Tailcuts for all of the studied Cherenkov cameras. We observe that more details are retained after clustering cleaning. Both methods demonstrate strong robustness against NSB, broken or unavailable pixels, and calibration uncertainties using simulations. However, tests on real data are necessary to compare and validate these results. Additionally, simulations of the full MST subarray equipped with FlashCam demonstrated improved angular and energy resolution with the Time-based Clustering algorithm compared to Tailcuts.

While refinements such as multi-peak identification in waveforms and weighted~(or 4D) cluster finding are of interest, initial studies suggest they offer minimal advantages in volume reduction, with the added cost of increased complexity (cf.\ Sec.~\ref{appendix}). 

Furthermore, this method is highly general, making it not only applicable to all cameras of the CTAO, where it demonstrates excellent performance, but also suitable for use by any other array of Cherenkov Telescopes as a {\it Data Volume Reduction} or {\it Image Cleaning} algorithm.  

\section*{CRediT authorship contribution}

\textbf{Clara Escañuela Nieves:} Writing - Original Draft, Writing - Review and Editing, Software, Validation, Visualization, Methodology, Investigation, Conceptualization, Analysis. \textbf{Felix Werner:} Writing - Review and Editing, Conceptualization, Supervision. \textbf{Jim Hinton:} Writing - Review and Editing, Supervision, Project administration.

\section*{Acknowledgements}

We gratefully acknowledge the use of {\it ctapipe} for validating the method. Our thanks go to Georg Schwefer for providing supporting tools for this validation. Clara Escañuela Nieves is a fellow of the International Max Planck Research School for Astronomy and Cosmic Physics at the University of Heidelberg~(IMPRS-HD).

\appendix
\section{Revised approaches to Time-based Clustering} \label{appendix}

Three modifications to the Clustering algorithm are studied and tested:

\begin{enumerate}
    \item Four-dimensional clustering, incorporating a fourth dimension related to the extracted charge, enhances the significance of each pixel's importance based on its charge;
    \item each point is weighted using a charge-dependent expression to increase the impact of points with higher charge and to favour the detection of small clusters;
    \item we explore adjustments to the signal extraction algorithm to refine the accuracy of charge and time reconstruction within each pixel.
\end{enumerate}

\begin{figure}
    \centering
    \includegraphics[width=\linewidth]{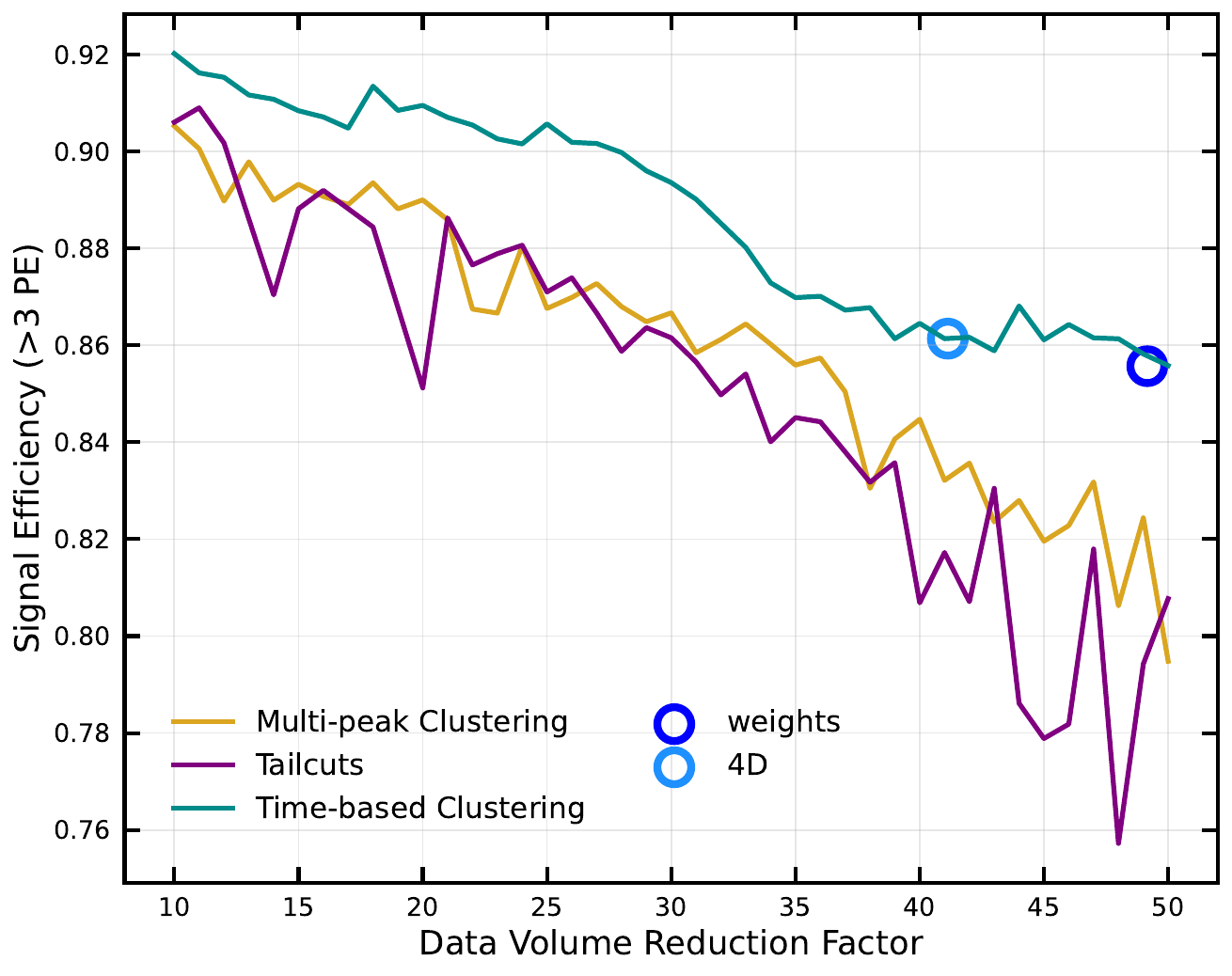}
    \caption{Signal efficiency for detecting pixels with more than 3\,PEs versus {\it Data Volume Reduction} factor. The default clustering and Tailcuts methods are compared to a clustering algorithm fed with more than one time value per waveform (Multi-peak clustering). Light blue: 4D clustering. Dark blue: weighted clustering. Both dots are plotted with the same parameters as the standard method at a DVR factor of 40. }
    \label{fig:sig_weights}
\end{figure}

Fig.~\ref{fig:sig_weights} shows the signal efficiency of pixels with more than three true PEs as a function of the DVR factor using proton simulations. It shows the evolution of signal efficiency with the DVR factor for the standard, previously tested, Time-based Clustering, Tailcuts, as well as the three methods mentioned above and that will be explained in detail below.

Firstly, we explore the addition of one extra dimension to the clustering based on the charge in that pixel. So far, only three dimensions have been considered, disregarding any charge information at the step when DBSCAN is applied. We test the effect of including a fourth dimension into DBSCAN clustering. Clustering the charge values is not feasible due to the lack of proximity between number of PEs of neighbour signal pixels. Therefore, the fourth dimension is defined as \(\log_{10}(\frac{\text{Q}\ /\ \text{PE}}{\text{Q}_{\text{max}}\ /\ \text{PE}})\), with \(\text{Q}_{\text{max}}\) being the maximum charge per-event. We now have a scenario where high charge values in the shower appear closer than noise pixels with low charge. The light blue point in Fig.~\ref{fig:sig_weights} shows the signal efficiency and DVR factor of the 4-dimensional method with the same free parameters of the standard Clustering method at a DVR of~40. No significant effect can be seen.

Alternatively, a continuous charge-dependent weighting is applied to the pixels to take into account their significance in amplitude. DBSCAN loops over every data point drawing a circular line around it with radius, eps. The number of points inside that area is compared to the minimum threshold, minPts. At this point of the algorithm, weights are summed inside the area instead of discrete points. Therefore, points with larger charge have a larger impact. The weighting factors have the following shape:

\begin{equation}
\text{weights} = \frac{1}{1\ +\ \exp{\frac{-(\text{PE}\ +\ A)}{B}}}
\end{equation}

This introduces two extra free parameters for optimization, $A$ and $B$. These parameters have been chosen as $2.0$ and $4.0$, respectively. Different forms for the weighting can be explored to improve the detection of very small clusters. The dark blue point in Fig.~\ref{fig:sig_weights} represents the signal efficiency and DVR factor of the weighting method with parameters choice from Table~\ref{tab:clust_parameters} at a DVR factor of 40. As shown, signal detection is lower compared to unweighted clustering, since reaching minPts is more challenging when weights (\(\leq 1\)) are applied. To achieve the same signal efficiencies, minPts should be reduced.

Charge and time extraction is crucial in any DVR/cleaning algorithm, as pixel-wise methods rely heavily on the accuracy of this extraction. To enhance the performance of clustering, additional signal extraction algorithms can be explored. By default, we use a method that yields a single signal charge and time value for each pixel. However, waveforms can be complex. PEs from showers do not all reach a pixel simultaneously, typically causing a time spread of about 2\,ns within each pixel. Additionally, signal pulses can be confused with significant noise pulses, such as afterpulses caused by NSB photons, leading to deviations in the reconstructed time within the bounds of the waveform (with the probability of deviation proportional to the waveform duration). These factors, among others, can result in multiple significant peaks in waveforms, as illustrated in Fig.~\ref{fig:wfs_pz}.

Traces, as shown in Fig.~\ref{fig:wfs_pz}, are a result of a convolution of multiple PEs pulses that arrive at different times within the waveform. The time of arrival of those PEs is given by the position of the peak of the multiple pulses in the differentiated trace. As we do not know which peak has information of the signal PEs, we can extract the time of each local maximum. The charge is given by the value in PE of the local maxima. Only maxima with amplitudes above a threshold are fed into the DBSCAN algorithm. Every pixel could have more than one reconstructed time value. Consequently, more points are typically also fed into DBSCAN, which slightly increases the computational time of the algorithm and also the number of noise points. This has an advantage for identifying multiple (sub)showers overlapping on the camera plane, e.g.\ muon rings on top of proton shower images. Those pixels sharing light from both events can now discriminate between the arrival time of PE from the proton shower and the muon. This can have applications in muon trigger/tagging and helps identifying multiple clusters in time.

The second step of the default algorithm relies on also keeping highly significant pixels that may have been missed initially. This additional step aims to ensure high signal efficiency for pixels with high charge, yet it proves less effective with the new method. This is because the maximum as a charge indicator is not as stable as the extracted charge. Therefore, this step is skipped on this alternative. Fig.~\ref{fig:sig_weights} illustrates the efficiency of detecting signal pixels. The algorithm described above (Multi-Peak Clustering, in the plot) demonstrates similar performance to Tailcuts.

The Multi-Peak Clustering method is very promising. It does need certain changes to solve two main problems. First, the detection of high significant pixels does not reach the efficiencies of Tailcuts or Clustering. Second, the method is very robust against calibration uncertainties but not with broken pixels. Broken pixels have a very problematic effect on this method as invalid pixels have so far zero intensity. This can cause gaps in the image which make the detection of clusters based on proximity harder. This could be improved by following the example of the standard method, Section~\ref{sec:broken}, and substitute invalid pixels by the average waveform of their nearest neighbours.

\section{CTAO South sub-array MSTs}\label{appendix_layout}

\begin{figure}[h!]
    \centering
    \includegraphics[width=\linewidth]{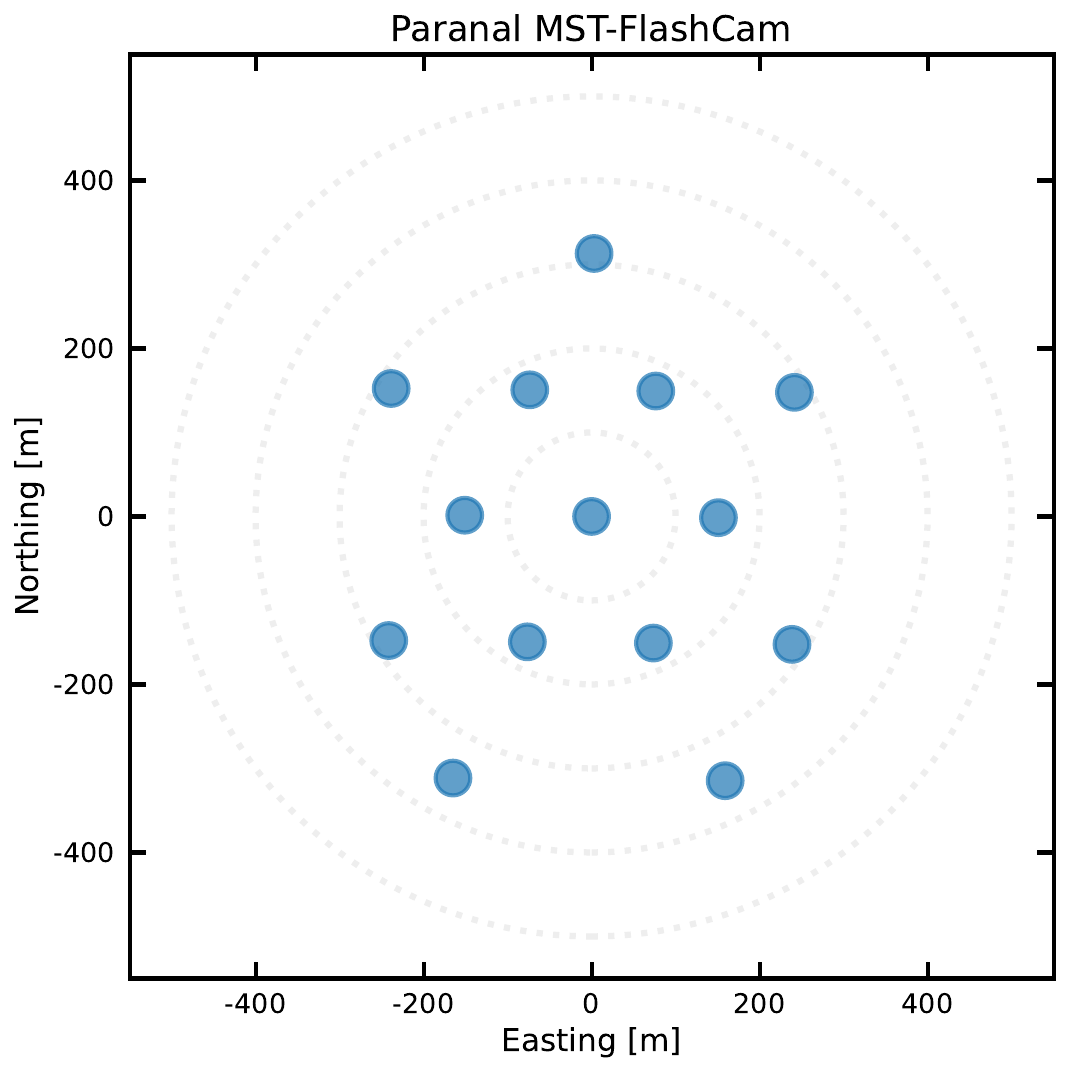}
    \caption{Layout of MST-FlashCam in the South~\citep{website_layout} with the simulated positions of the telescopes in the x and y directions. }
    \label{fig:alp_layout}
\end{figure}

Fig.~\ref{fig:alp_layout} shows the telescope layout of 14\,MSTs, equipped with FlashCam cameras, in the Southern array in Chile. This layout has been used to simulate the events used previous to make verification plots with FlashCam, including angular and energy resolution plots. 

\bibliography{references.bib}

\begin{thebibliography}{10}
\expandafter\ifx\csname url\endcsname\relax
  \def\url#1{\texttt{#1}}\fi
\expandafter\ifx\csname urlprefix\endcsname\relax\def\urlprefix{URL }\fi
\expandafter\ifx\csname href\endcsname\relax
  \def\href#1#2{#2} \def\path#1{#1}\fi

\bibitem{technique}
T.~C. Weekes, {The atmospheric Cherenkov technique in very high energy gamma-ray astronomy}, Space Science Reviews 75~(1) (1996) 1--15.

\bibitem{ctao}
S.~Vercellone, C.~Consortium, et~al., {The next generation Cherenkov Telescope Array observatory: CTA}, Nuclear Instruments and Methods in Physics Research Section A: Accelerators, Spectrometers, Detectors and Associated Equipment 766 (2014) 73--77.

\bibitem{website_layout}
{CTAO Website,} layouts for alpha configuration, \url{https://www.ctao.org/news/ctao-releases-layouts-for-alpha-configuration/}, (Accessed 08 November 2024).

\bibitem{website}
{CTAO Website,} data and computing, \url{https://www.ctao.org/emission-to-discovery/data-and-computing/}, (Accessed 08 November 2024).

\bibitem{flashcam}
F.~Werner, C.~Bauer, S.~Bernhard, M.~Capasso, S.~Diebold, F.~Eisenkolb, S.~Eschbach, D.~Florin, C.~F{\"o}hr, S.~Funk, et~al., {Performance verification of the FlashCam prototype camera for the Cherenkov Telescope Array}, Nuclear Instruments and Methods in Physics Research Section A: Accelerators, Spectrometers, Detectors and Associated Equipment 876 (2017) 31--34.

\bibitem{data_model_ref}
J.~L. Contreras, K.~Satalecka, K.~Bernl{\"o}hr, C.~Boisson, J.~Bregeon, A.~Bulgarelli, G.~de~Cesare, R.~Reyes, V.~Fioretti, K.~Kosack, et~al., {Data model issues in the Cherenkov Telescope Array project}, arXiv preprint arXiv:1508.07584 (2015).

\bibitem{hillas}
A.~M. Hillas, {Cerenkov light images of EAS produced by primary gamma}, in: 19th Intern. Cosmic Ray Conf-Vol. 3, no. OG-9.5-3, 1985.

\bibitem{freepact}
G.~Schwefer, R.~Parsons, J.~Hinton, {A hybrid approach to event reconstruction for atmospheric Cherenkov Telescopes combining machine learning and likelihood fitting}, Astroparticle Physics 163 (2024) 103008.

\bibitem{pz_dec}
C.~Nowlin, J.~Blankenship, {Elimination of undesirable undershoot in the operation and testing of nuclear pulse amplifiers}, Review of Scientific Instruments 36~(12) (1965) 1830--1839.

\bibitem{tailcuts}
M.~Punch, C.~Akerlof, M.~Cawley, D.~Fegan, R.~Lamb, M.~Lawrence, M.~Lang, D.~Lewis, D.~Meyer, K.~O'Flaherty, et~al., {Supercuts: an improved method of selecting gamma-rays}, in: Proceedings of the 22nd International Cosmic Ray Conference. 11-23 August, 1991. Dublin, Ireland. Under the auspices of the International Union of Pure and Applied Physics (IUPAP), Volume 1, Contributed Papers, OG Sessions 1-5. Dublin: The Institute for Advanced Studies, 1991., p. 464, Vol.~1, 1991, p. 464.

\bibitem{ctapipe}
M.~Linhoff, S.~Bhattacharyya, J.~P{\'e}rez~Romero, S.~Stani{\v{c}}, V.~Vodeb, S.~Vorobiov, D.~Zavrtanik, M.~Zavrtanik, M.~{\v{Z}}ivec, {ctapipe-prototype open event reconstruction pipeline for the Cherenkov Telescope Array}, in: 38th International Cosmic Ray Conference [also] ICRC2023, 2023, pp. 1--14.

\bibitem{impact}
R.~Parsons, J.~Hinton, {A Monte Carlo template based analysis for air-Cherenkov arrays}, Astroparticle Physics 56 (2014) 26--34.

\bibitem{dbscan}
M.~Ester, H.-P. Kriegel, J.~Sander, X.~Xu, et~al., {A density-based algorithm for discovering clusters in large spatial databases with noise}, in: kdd, Vol.~96, 1996, pp. 226--231.

\bibitem{hdbscan}
R.~J. Campello, D.~Moulavi, J.~Sander, {Density-based clustering based on hierarchical density estimates}, in: Pacific-Asia conference on knowledge discovery and data mining, Springer, 2013, pp. 160--172.

\bibitem{optics}
M.~Ankerst, M.~M. Breunig, H.-P. Kriegel, J.~Sander, {OPTICS: Ordering points to identify the clustering structure}, ACM Sigmod record 28~(2) (1999) 49--60.

\bibitem{time-clustering}
S.~Steinma{\ss}l, {Probing particle acceleration in stellar binary systems using gamma-ray observations}, Ph.D. thesis, Ruprecht-Karls-Universit{\"a}t Heidelberg (2023).

\bibitem{corsika}
D.~Heck, J.~Knapp, J.~Capdevielle, G.~Schatz, T.~Thouw, et~al., {CORSIKA: A Monte Carlo code to simulate extensive air showers} (1998).

\bibitem{simtelarray}
K.~Bernl{\"o}hr, {Simulation of imaging atmospheric Cherenkov telescopes with CORSIKA and sim\_telarray}, Astroparticle Physics 30~(3) (2008) 149--158.

\bibitem{alpha-config}
R.~Zanin, {CTA--the World’s largest ground-based gamma-ray observatory}, in: Proceedings of 37th International Cosmic Ray Conference (ICRC2021), Vol. 395, SISSA, 2022, p. 005.

\bibitem{aharonian1997potshowers}
F.~A. Aharonian, W.~Hofmann, A.~Konopelko, H.~V{\"o}lk, {The potential of ground based arrays of imaging atmospheric Cherenkov telescopes. I. Determination of shower parameters}, Astroparticle Physics 6~(3-4) (1997) 343--368.

\bibitem{ang_rec}
F.~A. Aharonian, W.~Hofmann, A.~Konopelko, H.~V{\"o}lk, {The potential of ground based arrays of imaging atmospheric Cherenkov telescopes. I. Determination of shower parameters}, Astroparticle Physics 6~(3-4) (1997) 343--368.

\bibitem{gammapy}
A.~Donath, C.~Deil, M.~P. Arribas, J.~King, E.~Owen, R.~Terrier, I.~Reichardt, J.~Harris, R.~B{\"u}hler, S.~Klepser, {Gammapy-A Python package for gamma-ray astronomy}, arXiv preprint arXiv:1509.07408 (2015).

\bibitem{python}
G.~Van~Rossum, F.~L. Drake, et~al., {Python reference manual}, Vol. 111, Centrum voor Wiskunde en Informatica Amsterdam, 1995.

\bibitem{bright-stars}
R.~Cornils, S.~Gillessen, I.~Jung, W.~Hofmann, M.~Beilicke, K.~Bernl{\"o}hr, O.~Carrol, S.~Elfahem, G.~Heinzelmann, G.~Hermann, et~al., {The optical system of the HESS imaging atmospheric Cherenkov telescopes. Part II: mirror alignment and point spread function}, Astroparticle Physics 20~(2) (2003) 129--143.

\bibitem{sst}
D.~Depaoli, {Status of the SST camera for the Cherenkov Telescope Array}, arXiv preprint arXiv:2310.01183 (2023).

\bibitem{nectarcam}
T.~Tavernier, J.-F. Glicenstein, F.~Brun, {Status and performance results from NectarCAM--a camera for CTA medium sized telescopes}, arXiv preprint arXiv:1909.01969 (2019).

\bibitem{lst}
Y.~Kobayashi, A.~Okumura, F.~Cassol, H.~Katagiri, J.~Sitarek, P.~Gliwny, S.~Nozaki, Y.~Nogami, {Camera Calibration of the CTA-LST prototype}, arXiv preprint arXiv:2108.05035 (2021).

\end{thebibliography}
\bibliographystyle{elsarticle-num} 

\end{document}